\documentclass[
 reprint,
superscriptaddress,
 amsmath,amssymb,
 aps,
prd,
floatfix,
]{revtex4-2}
\usepackage[dvipsnames]{xcolor}
\usepackage{graphicx}%
\usepackage{dcolumn}%
\usepackage{bm}%
\usepackage{siunitx}
\usepackage{xcolor}
\usepackage{xspace}
\usepackage{booktabs}
\usepackage{multirow}
\usepackage{physics}
\usepackage{makecell}
\usepackage[breaklinks=true]{hyperref}
\usepackage[normalem]{ulem}
\usepackage{cleveref}

\usepackage{afterpage}

\newcommand{\gibuu}{\textsc{GiBUU}\xspace}
\newcommand{\nuwro}{\textsc{NuWro}\xspace}
\newcommand{\minerva}{MINERvA\xspace}
\newcommand{\nova}{NOvA\xspace}
\newcommand{\microboone}{MicroBooNE\xspace}

\newcommand{\dat}{\delta\alpha_\textrm{T}}
\newcommand{\pn}{p_\textrm{N}}
\newcommand{\Tpar}{\mathcal{T}}
\newcommand{\Ttpth}{\Tpar_2}
\newcommand{\Toneb}{\Tpar_1}
\newcommand{\pythia}{\textsc{Pythia}\xspace}
\newcommand{\twopibg}{2$\pi$BG\xspace}
\newcommand{\ubang}{\cos \theta_{\pi^0}}
\newcommand{\ubmom}{p_{\pi^0}}
\newcommand{\lprime}{{\ell^\prime}}
\newcommand{\nprime}{{\textrm{N}^\prime}}
\newcommand{\figref}[1]{Fig.~\ref{#1}}
\newcommand{\gibver}{25p1}

\newcommand{\widefigwid}{0.49\textwidth}
\newcommand{\figwid}{0.43\textwidth}
\definecolor{darkgreen}{rgb}{0.0, 0.5, 0.0}
\definecolor{darkblue}{rgb}{0.0, 0.0, 0.7}

\newcommand{\warwick}{University of Warwick, Department of Physics, Coventry, CV4 7AL United Kingdom}
\newcommand{\ucas}{School of Physical Sciences, University of Chinese Academy of Sciences, Beijing 100049, China}
\newcommand{\giessen}{Institut f\"ur Theoretische Physik, Universit\"at Giessen, 35392 Giessen, Germany}

\newcommand{\ihep}{Institute of High Energy Physics, Beijing, China}

\crefname{section}{Sec.}{Secs.}
\crefname{subsection}{Sec.}{Secs.}
\crefname{figure}{Fig.}{Figs.}
\crefname{table}{Table}{Tables}
\crefname{equation}{Eq.}{Eqs.}
\creflabelformat{equation}{#2#1#3}

\begin{document}
\title{Understanding neutrino pion production with the \gibuu model}%

\author{Qiyu Yan}
\email{yanqiyu17@mails.ucas.edu.cn}
\affiliation{\ucas}%
\affiliation{\warwick}%

\author{Kaile Wen}
\affiliation{\ihep}

\author{Kai Gallmeister}
\email{kai.gallmeister@gmail.com}
\affiliation{\giessen}

\author{Xianguo Lu}
\email{xianguo.lu@warwick.ac.uk}
\affiliation{\warwick}

\author{Ulrich Mosel}
\email{mosel@physik.uni-giessen.de}
\affiliation{\giessen}

\author{Yangheng Zheng}
\affiliation{\ucas}

\date{\today}%

\begin{abstract}

    Pion production is a major source of systematic uncertainty in neutrino oscillation measurements. We report a systematic investigation of neutrino-induced pion production using \minerva and \microboone data within the \gibuu theoretical framework. The analysis begins by establishing baseline model parameters using inclusive and pionless data from \minerva, \microboone, and T2K experiments. We then examine the role of in-medium effects, including resonance broadening and nucleon-nucleon final-state interactions. While agreement with individual datasets can be achieved through specific model configurations, we demonstrate the difficulty of a unified description across all experiments: \minerva measurements prefer  minimum in-medium modifications, whereas \microboone data require the maximum in-medium enhancement, revealing the complexity and richness of the underlying nuclear dynamics.
\end{abstract}

\maketitle

\section{Introduction}\label{sec:intro}

Neutrinos serve as powerful probes of the fundamental properties of the universe.
Experiments such as DUNE~\cite{DUNE:whitepaper}, Hyper-Kamiokande~\cite{HK}, NOvA~\cite{nova}, and T2K~\cite{ABE2011106} aim to measure neutrino oscillations to investigate how neutrino flavor mixing violates charge-parity (CP) symmetry.
In addition, DUNE aims to determine the neutrino mass hierarchy, which is another key question in neutrino physics~\cite{abi2020deep}.
In those measurements, understanding the pion production process is essential for accurately  determining the oscillated flux.
However, pion production remains poorly understood.
In response, there have been efforts in both tuning model parameters to better align with data~\cite{PhysRevD.105.012009,PhysRevD.110.072016} and developing new theoretical frameworks~\cite{Yan:2024kkg}.

The Giessen-Uehling-Uhlenbeck (GiBUU) is a common theoretical and numerical framework for the description of nuclear collisions. Its applicability ranges from relativistic heavy-ion collisions to hadron + nucleus and photon-, electron-, and neutrino-induced collisions, and it has been widely tested against all these very different reaction types. Common to all these different reactions is the description of final-state-interactions which do not depend on the particular reaction type but just on the initially deposited energy and momentum. Its particular strength is its description of final state interactions by using quantum-kinetic transport theory; this makes it unique among all the generators used for photon, electron and neutrino reactions. A comprehensive description of the theory and other model ingredients, including their numerical implementation, can be fond in Ref.~\cite{Buss:2011mx}; the code is freely available and can be obtained from Ref.~\cite{gibuu} where also many help texts can be found.

Recently, there have been significant efforts by \gibuu to improve the description of pion production, including, in particular, that of the $2\pi$-background contribution~\cite{Mosel:2023zek}. This process involves the production of two pions without passing through intermediate resonance states.
The earlier work of Ref.~\cite{Mosel:2023zek} was restricted to semi-inclusive cross sections. However, a detailed comparison to demonstrate the impact of these background contributions against more exclusive experimental data on pion production is still lacking.
Apart from a detailed study in Ref.~\cite{bogart2024inmedium} on medium effects in $^{40}\text{Ar}$ neutral-current interactions---which found experimental preference for including in-medium  cross section modifications---a thorough model-data comparison  remains crucial for understanding its impact on pion production.
This work aims to fill this gap by comparing the most up-to-date \gibuu (version \gibver) predictions with multiple experimental results, focusing on the influence and potential sources of systematic uncertainties introduced by these new background contributions.

In the following, the \gibuu model will be reviewed in Sec.~\ref{sec:gibuu}. In Sec.~\ref{sec:tki}, the transverse kinematic imbalance (TKI) framework, one of the key  observables to be compared between models and data, will be described.
In Secs.~\ref{sec:cc0pi}--~\ref{sec:systematics}, detailed comparison will be made to address the impact of the new background contributions in pionless and pion productions and the relevant sources of systematics. A summary is presented in Sec.~\ref{sec:summary}.

\section{The \gibuu Model}\label{sec:gibuu}
In \gibuu the initial interaction and the final state interactions do not factorize since both take place in the same nuclear potential.
For neutrino-nucleus interactions the initial processes are quasielastic scattering (QE), $\Delta$ resonance and higher resonance excitations, and deep inelastic scattering (DIS). These are all one-body processes which also happen in a neutrino-nucleon reaction. Details about their description can be found in Refs.~\cite{Leitner:2006ww,Lalakulich:2012gm}. In a neutrino-nucleus reaction the so-called 2p2h (or meson exchange current---MEC) also appear; their implementation is described in  Ref.~\cite{Mosel:2023zek}.

The initial-and final-state particles participating a neutrino interaction are placed within a mean-field potential~\cite{Buss:2011mx,Mosel:2023zek} which, for simplicity, is assumed to be a Lorentz-scalar. The final-state interactions (FSIs) manifest as the final-state particles'  subsequent transport out of the nucleus. The off-shellness of bound particles that are subjected to this potential is accounted for and evolves with the potential as they propagate. Essential is further that this potential is not only position-, but also momentum-dependent. The former requires a numerical integration of each particle's trajectory, the latter complicates energy-momentum conservation in final state secondary collisions \cite{Gallmeister:2025jwe}.

\subsection{Pion Production} \label{subsec:pion}
In \gibuu, single-pion production is dominated by the $\Delta$-resonance at low hadronic invariant mass, $W$, in (anti)neutrino interactions with nucleons. Higher resonances also contribute significantly to $\nu\textrm{n}$ and $\bar{\nu}\textrm{p}$ interactions beyond the $\Delta$-resonance region.

The \gibuu resonance contributions are implemented by including resonances up to \SI{2}{GeV} in mass.
Their vector form factors follow the MAID 2007 analysis~\cite{MAID07}, except for the $\Delta$-resonance: Since $\Delta$ production is more tightly constrained by pion production data, a specialized form factor is used for that case~\cite{Leitner:2006ww,Lalakulich:2010ss,Mosel:2023zek}. While MAID extensively covers single-pion production, it omits multi-pion processes. For (non-resonant) background contributions in the resonance region, \gibuu assumes that 1$\pi$ or 2$\pi$ final states dominate~\cite{Mosel:2023zek}. They are referred to as $1\pi$BG and $2\pi$BG. The neutrino cross sections for these background terms  are obtained from the Bosted-Christy parametrization~\cite{Christy:2007ve, Bosted:2007xd}, which is based on fits to electron-nucleon scattering data up to $W = \SI{3}{GeV}$ and four-momentum-transfer-squared $Q^2 = \SI{8}{GeV^2}$. The electron cross sections are converted into those for neutrinos by using Eq.~\ref{structfunct}.

DIS sets in at $ W = 2$ GeV. In the so-called Shallow Inelastic Scattering (SIS) region between 2 GeV and 3 GeV the cross sections
are modeled using again the Bosted-Christy parametrization. The actual final state is, however, generated by \pythia.

At higher invariant masses ($W > \SI{3}{GeV}$), \gibuu relies exclusively on \pythia to model the cross sections and final-state products. These events are classified as “true DIS”, in contrast to low-$W$ DIS events with $2 < W < \SI{3}{GeV}$, where the cross section is derived from the Bosted-Christy parametrization but the final-state generation is handled by \pythia.

\subsubsection{Pion background contributions}\label{subsec:Tpar}
The neutrino structure functions for background contributions are obtained from the electron counterpart via the following relations between the neutrino and electron structure functions, $W_{1,3}^\nu$ and $W_{1,3}^e$, respectively. These relations assume that the given processes are transverse which is indeed to a good accuracy the case. The electron structure functions for the background contributions are obtained from an analysis of electron scattering data in a wide kinematical range \cite{Christy:2007ve,Bosted:2007xd}. The cross section is then evaluated first in the nucleon rest frame using Eq.\ (1) in \cite{Mosel:2023zek}. The arguments of the structure functions, i.e.\ $Q^2$ and the energy transfer $\omega$, are taken to be those for a nucleon at rest \cite{Mosel:2023zek}; the cross sections are then boosted from the nucleon rest frame to the nucleus rest frame in order to account for Fermi motion \cite{Mosel:2023zek}.

The neutrino structure functions are then given by
\begin{align}   \label{structfunct}
    W_1^\nu & = \left[1 + \left( \frac{2m}{\mathbf{q}} \right)^2  \left( \frac{G_A(Q^2)}{G_M(Q^2)} \right)^2 \right] 2(\Tpar +1)W_1^e, \\
    W_3^\nu & = 2 \left( \frac{2m}{\mathbf{q}} \right)^2   \frac{G_A(Q^2)}{G_M(Q^2)} \, 2(\Tpar +1)W_1^e.
\end{align}
Here, $m$ is the nucleon mass, $q$ is the momentum transfer magnitude, and $G_A(Q^2)$ and $G_M(Q^2)$ are the weak axial and electromagnetic isovector form factors.

The justification for these connections between the electron- and the neutrino-structure-functions has been discussed in Sec. III C in Ref.~\cite{Mosel:2023zek}. Here we just mention that the pion background contributions arise mostly from nucleon-pion processes (all resonance contributions to pion production are treated explicitly in GiBUU~\cite{Leitner:2006ww}); this can be seen, e.g., in the work of Hernandez \textit{et al}.~\cite{Hernandez:2007ej}. The kinematical factor $2m/\mathbf{q}$ in Eqs.\ (1) and (2) then just reflects the factor that appears in the transition matrix elements of free nucleons~\cite{walecka2012semileptonic}. The form factors appearing there are taken, for simplicity, to be those of the $W$-boson-nucleon vertex, thus preserving current conservation~\cite{Hernandez:2007ej}.

In the original derivation by Walecka \textit{et al.}~\cite{walecka2012semileptonic,OConnell:1972edu}, the quantity $\Tpar$ is the isospin of the nucleus. This derivation is based on a single-particle model even for high excitations and assumes that the final states seen in electron scattering are the isobaric analogues of those seen in neutrino interactions. Both of these assumptions are doubtful; therefore, in the present version of GiBUU, $\Tpar$ is allowed to vary in a reasonable range.

The parameterizations of background cross sections given in \cite{Christy:2007ve,Bosted:2007xd} do not contain any information on specific final states. In GiBUU the background cross sections are split into 1$\pi$ and 2$\pi$ final states by an adjustment to experimental data on electron-induced 2$\pi$ data. Furthermore,
when going to neutrino-induced reactions it is assumed that the cross sections for neutrino-induced reactions are distributed among the final $1\pi N$ or $2\pi N$ states as they are for electron-induced ones.

\subsection{2p2h Process}  \label{subsec:2p2h}
Furthermore, the same transformation (Eq.~\ref{structfunct}) is also used as an approximation for the 2p2h contribution to the structure functions (for details see Ref.~\cite{Mosel:2023zek}). Here the neutrino structure functions given above have been shown to give a good description of of 2p2h processes both for neutrino and antineutrino reactions~\cite{Gallmeister:2016dnq}, without any readjustment. In the recent version 2025 of \gibuu the scaling factor $\Tpar$ is handled separately for the one-body interactions and the 2p2h channel. In the following text, we refer to it as $\Toneb$ for one-body channels, $\Ttpth$ for 2p2h, and collectively as $\Tpar$.

\subsection{Medium Effects}\label{Medium}

Inside a nucleus, hadronic processes, such as resonance decay and nucleon-nucleon scattering, can differ from those in a vacuum. This is because the participating particles  interact with the surrounding nucleons. We will discuss \gibuu's in-medium modifications in the following.

\gibuu provides a description of $\Delta$-resonance broadening in a nuclear medium due to collisional effects~\cite{Oset:1987re, OsetSalcedo}. This so-called Oset broadening influences the resonance's effective width. The procedure is implemented by replacing the free width, $\Gamma_{\Delta}^\text{free}$, with its in-medium value, $\Gamma_{\Delta}^\text{medium}$:
\begin{equation}
    \Gamma_{\Delta}^\text{free} \rightarrow \Gamma_{\Delta}^\text{medium} = \Gamma_{\Delta}^\text{PB} - 2 \imaginary (\Sigma_{\Delta}).
\end{equation}
In this replacement, $\Gamma_{\Delta}^\text{PB}$ refers to the effect where the decay products of the $\Delta$-resonance may be Pauli-blocked in the nucleus, leading to a reduction in the resonance width. The term $\imaginary (\Sigma_{\Delta})$ represents the imaginary part of the $\Delta$ self-energy, which is associated with the collisional broadening of the resonance. This broadening arises from decay channels such as $\Delta \textrm{N} \rightarrow \pi \textrm{N}\textrm{N}$, $\Delta \textrm{N} \rightarrow \textrm{N}\textrm{N}$, and $\Delta \textrm{N}\textrm{N} \rightarrow \textrm{N}\textrm{N}\textrm{N}$, which are permitted only within the nuclear medium. The overall effect, which results from the interplay between Pauli blocking and collisional broadening, is an 'internal' broadening of the $\Delta$-resonance in the nucleus. Contrary to common belief, the main observable effect of this ``internal'' broadening is, however, only a suppression of the peak value of the cross-section; the broadening itself is hard to observe due to nuclear surface effects and the Fermi-motion \cite{Lehr:2001ju}.

In-medium  interactions are applied to FSI processes in \gibuu~\cite{Li:1993rwa,Li:1993ef}. The modification lowers the cross section of elastic  scattering as a function of density. In addition, the inelastic process related to the $\Delta$-resonance is also modified using the following description~\cite{SongKo}:
\begin{equation}   \label{Delta-supp}
    \sigma_{\textrm{N}\textrm{N}\rightarrow \textrm{N}\Delta}(\rho_\text{N})  = \sigma_{\textrm{N}\textrm{N}\rightarrow \textrm{N}\Delta}(0) \exp\left\{-\alpha \frac{\rho_\text{N}}{\rho_0}\right\},
\end{equation}
where $\rho_\text{N}$ is the nucleon density where the interaction occurs and $\alpha = 1.2$ is a fitting parameter. And as detailed balance applies to the reverse process, $\textrm{N}\Delta\rightarrow\textrm{N}\textrm{N}$, the corresponding cross section is also modified. As a result,  in neutrino-nucleus reactions in-medium changes will suppress both $\pi$ production and absorption during FSI.

The baseline model used in the following comparison to experimental data in Sec.~\ref{sec:pipro} is without the Oset broadening and is assuming free  interactions, which is the default setting in \gibuu 2025. Note that in the present calculations the $\Delta$ suppression of Eq.~\ref{Delta-supp} and the Oset broadening are linked together.  The effect of altering these medium effects will be discussed in  Sec.~\ref{sec:systematics}.

\subsection{Final State Evolution}

\begin{figure}[!tb]
    \centering
    \includegraphics[width=\widefigwid]{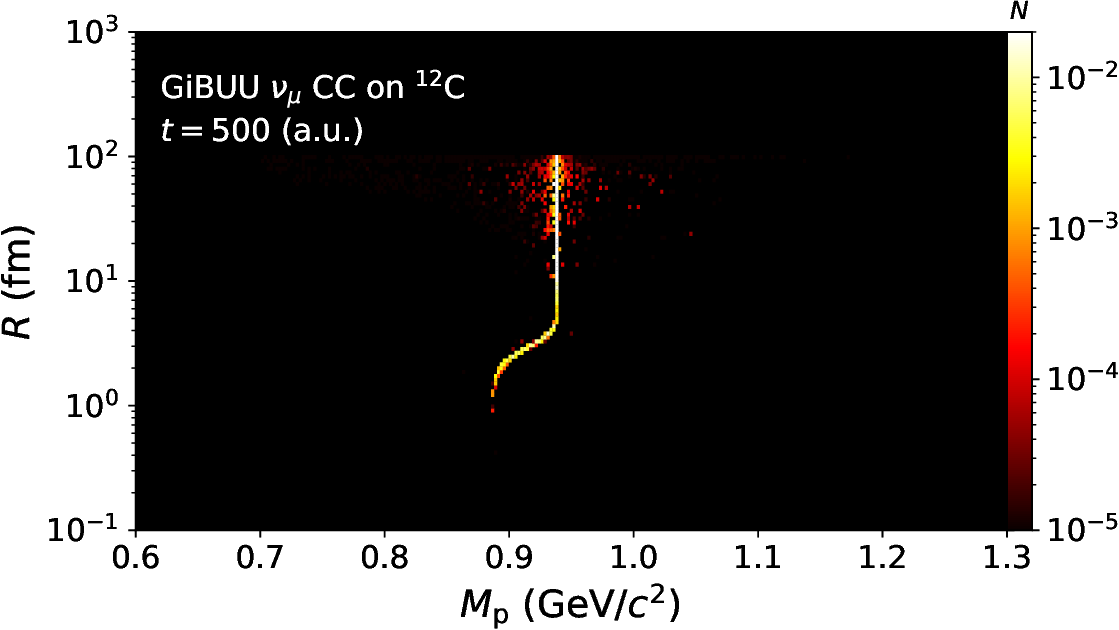} \\ \vspace{0.2cm}
    \includegraphics[width=\widefigwid]{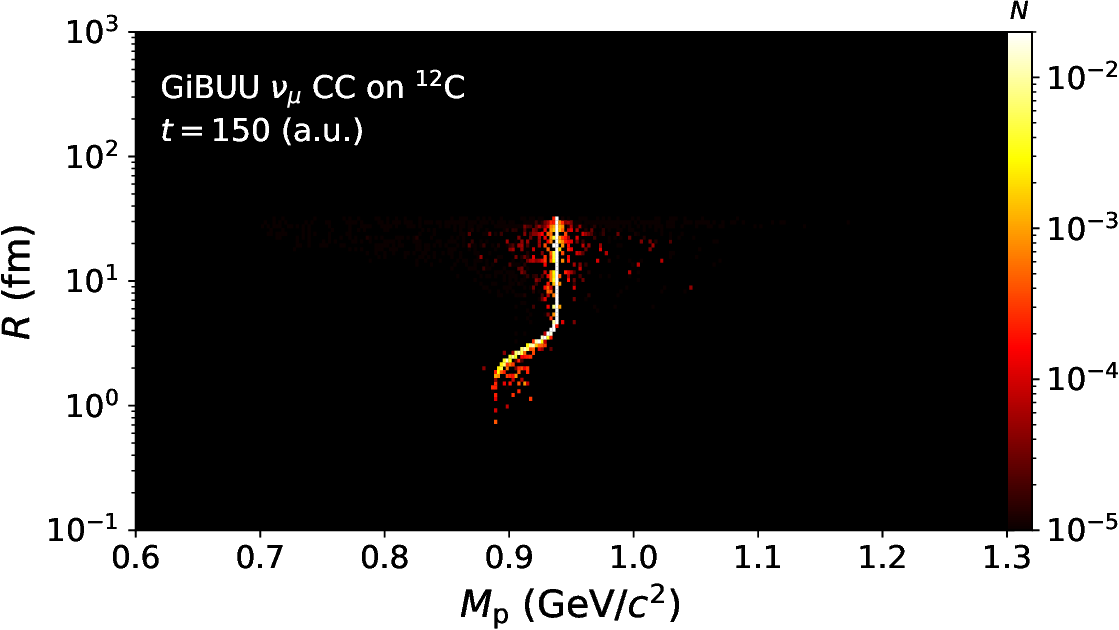} \\ \vspace{0.2cm}
    \includegraphics[width=\widefigwid]{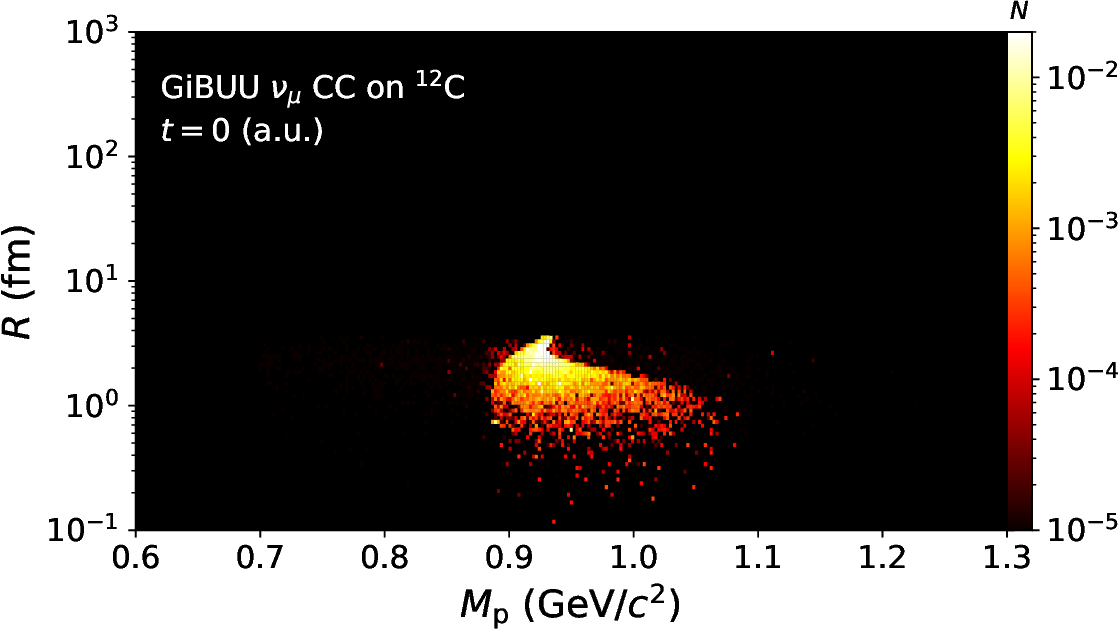}
    \caption{\gibuu $\nu_\mu$ CC interactions on $^{12}$C at time after interaction: $t=0,
            ~150,~\text{and}~500$ a.u. (from lower to upper), where $1~\text{a.u.}=0.2~\text{fm/}c\simeq 6.7\times10^{-25}~\textrm{s}$. The neutrino energy is 5~GeV. The population of the final-state protons ($N$ in an arbitrary unit) is shown as a function of the proton location, $R$, and its mass, $M_\textrm{p}$. The centre of the nucleus is at $R=0~\textrm{fm}$, and the boundary of the mean-field potential is seen at $R\simeq 5~\textrm{fm}$. The mass spread at large $R$ is a residual numerical artifact introduced during elastic scattering inside the nucleus.}     \label{fig:movie_m}
\end{figure}

\begin{figure}[!tb]
    \centering
    \includegraphics[width=\widefigwid]{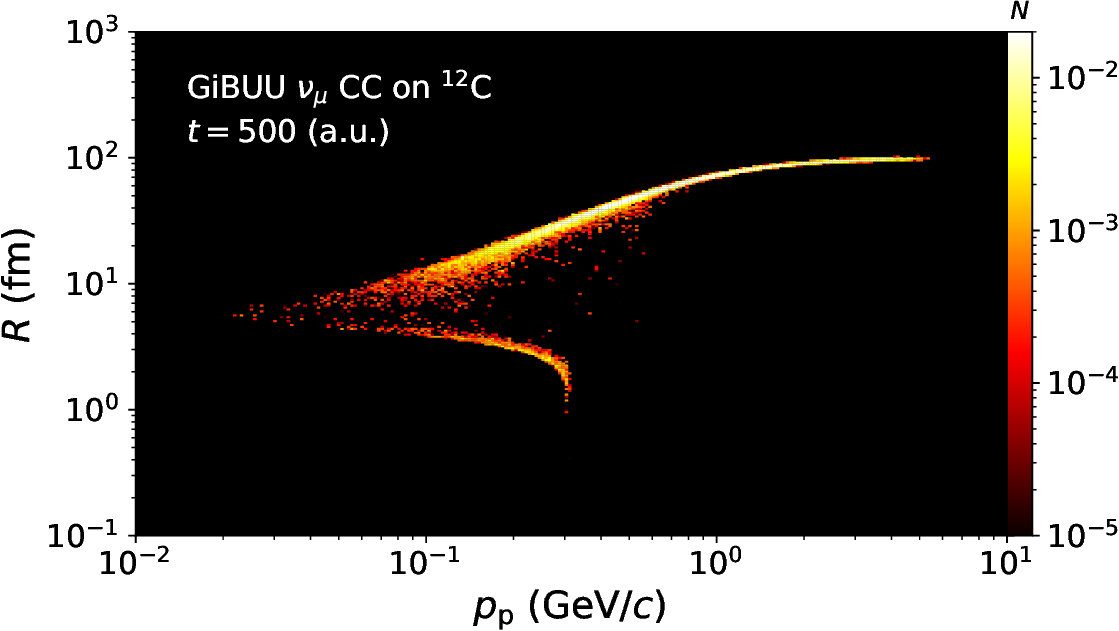} \\ \vspace{0.2cm}
    \includegraphics[width=\widefigwid]{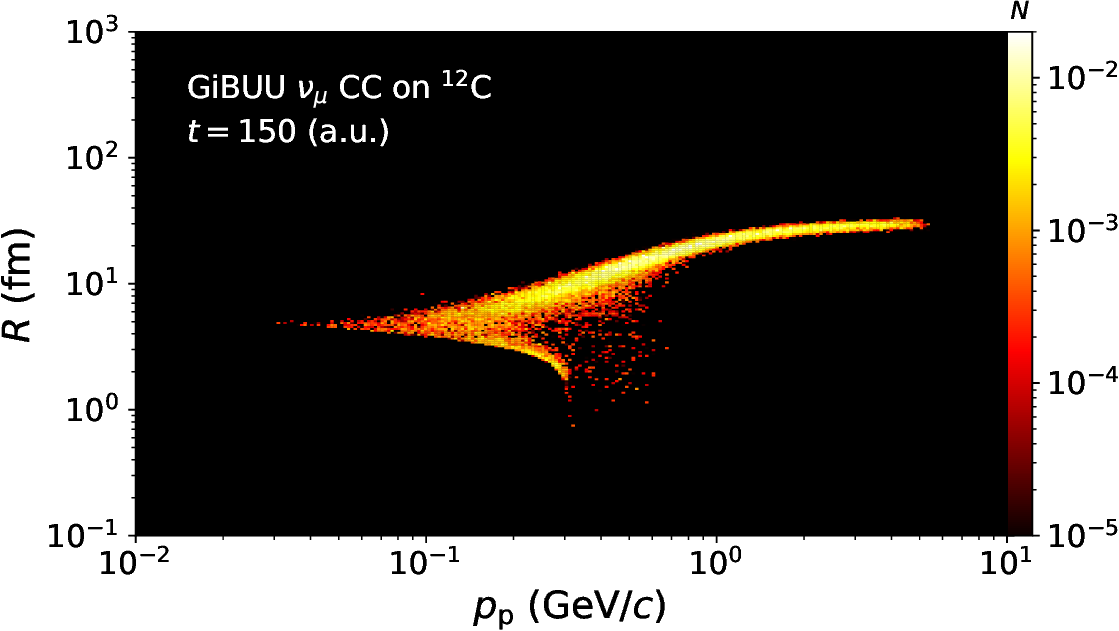} \\ \vspace{0.2cm}
    \includegraphics[width=\widefigwid]{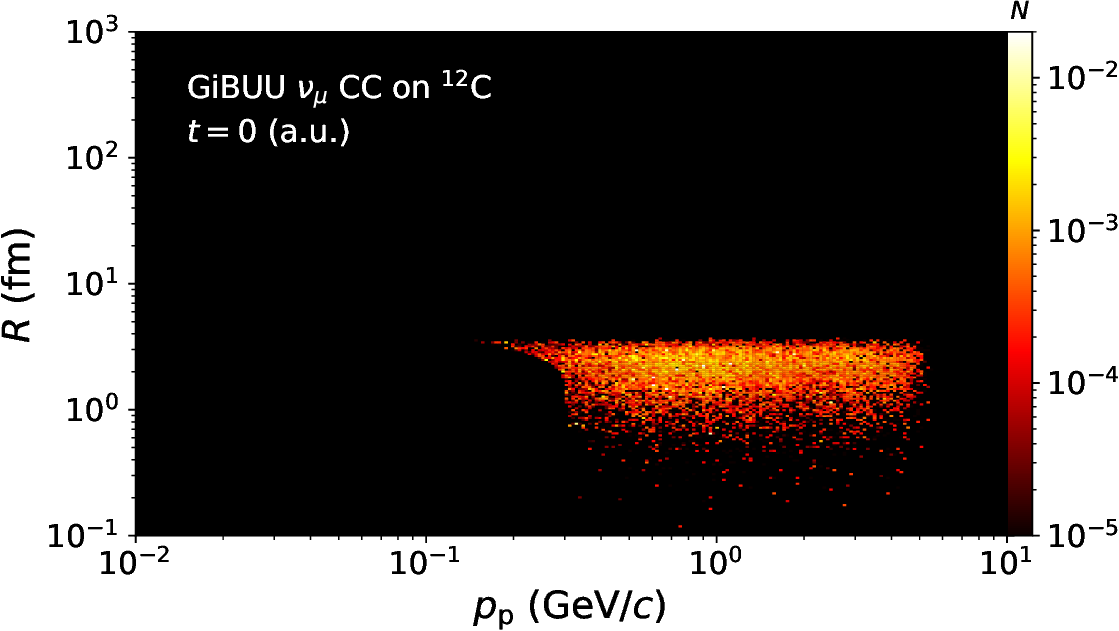}
    \caption{Similar to \figref{fig:movie_m} but as a function of the proton momentum $p_\textrm{p}$.}     \label{fig:movie_p}
\end{figure}

The evolution for final-state particles in \gibuu is illustrated in Figs.~\ref{fig:movie_m} and~\ref{fig:movie_p}~\footnote{In the GiBUU simulations, the final-state propagation time is set to different values, while the same random seed is used to generate the sample of GiBUU events. Whenever a proton hits another proton during the FSI stage both of these protons are propagated
    from then on. At the end of the propagation, the recorded positions and the four-momentum vectors of all final-state protons are analysed, and the proton mass $M_\textrm{p}$ is calculated as $\sqrt{E^2 - p^2}$, where $E$ and $p$ are the energy and momentum of the protons. The movies from which Figs.~\ref{fig:movie_m} and~\ref{fig:movie_p} are derived are available at~\url{https://luxianguo.github.io/html/gibuu/}.}. At time-zero following the interaction, the mass of the proton ranges between 0.88 and 1.05~GeV/$c^2$\footnote{The free proton mass in GiBUU is taken to be 0.938 GeV. Mass values below this free mass indicated binding.}. As the proton starts to propagate, its mass converges to a value dependent on its radial position within the nucleus. Once the proton exits the range of the mean-field potential---approximately 5~fm for carbon---it becomes on-shell.

Of greater experimental relevance is the evolution of the final-state momentum within the mean-field modeled by \gibuu. As shown in \figref{fig:movie_p}, the proton momentum initially exceeds the Fermi level which is approximately 0.2--0.3~GeV/$c$, and extends to the maximally allowed energy---the neutrino energy---at time zero. Protons that do not undergo intranuclear scattering retain their momentum throughout the transport. However, those that do scatter lose energy as they move outward towards the boundary of the mean-field potential, whereas their momenta are forbidden from falling below the Fermi level due to the Pauli exclusion principle. Consequently, in \figref{fig:movie_p}, a group of protons can be seen tracing the Fermi momentum, which decreases with increasing radial position due to the lower local density. Notably, some protons lose so much energy that they are unable to escape the potential, finding themselves bound within the target nucleus. On the one hand, future precision neutrino oscillation measurements rely on full-phase-space cross sections with final-state protons below  0.3~GeV/$c$, and are sensitive to such details~\cite{DUNE:whitepaper}. On the other hand, existing  experiments have a proton tracking threshold above 0.3~GeV/$c$ (see, e.g., Refs.~\cite{MicroBooNE:2020akw,T2K:2018rnz,MINERvA:2018hba,MINERvA:2021csy}), preventing protons at this low momentum from being observed.

\section{Transverse Kinematic Imbalance} \label{sec:tki}

The measurement of transverse kinematic imbalance (TKI)~\cite{Lu:2015hea,Lu:2015tcr} relies on the principle of momentum conservation in the transverse plane. By analyzing the discrepancy between the observed final state and the expected outcome, assuming the interaction occurs on a free nucleon target, the properties of the underlying nuclear model can be characterized.

\begin{figure}[!htb]
    \includegraphics[width=\figwid]{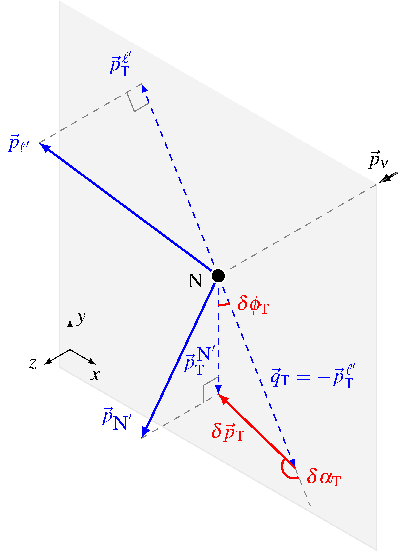}
    \caption{TKI schematic from Ref.~\cite{Lu:2015tcr}.
    }
    \label{fig:tkioriginal}
\end{figure}

\begin{figure}[!htb]
    \includegraphics[width=\figwid]{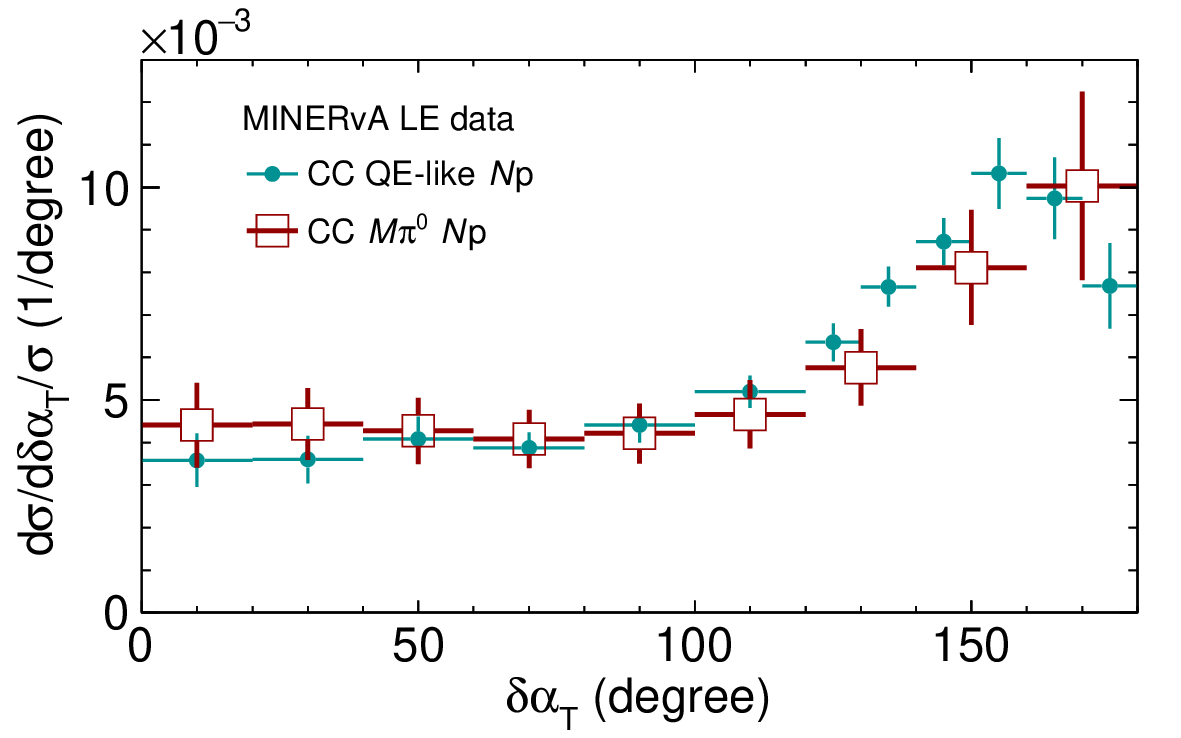}
    \caption{Shape-only comparison of $\dat$ from  CC0$\pi$~\cite{MINERvA:2018hba} and CC$\pi^0$ measurements~\cite{MINERvA:2020anu} by \minerva, where the final-state N is a proton and a $\pi+$proton system, respectively. Figure from Ref.~\cite{MINERvA:2020anu}. These data sets will be analyzed in Secs.~\ref{sec:cc0pi} and~\ref{subsec:ccpi0} where details will be given.
    }
    \label{fig:datcompile}
\end{figure}

Consider a reaction involving a neutrino and a nucleus:
\begin{equation}
    \nu + \text{A} \rightarrow \lprime + \text{X} + \text{N}^\prime
\end{equation}
where $\text{A}$ and $\text{X}$ refer to initial-state nucleus and final-state nuclear remnant, respectively, $\lprime$ is the final-state charged lepton, and $\text{N}^\prime$ is the nucleon knocked out of the nucleus in the pionless event, as proposed in Ref.~\cite{Lu:2015tcr} (Fig.~\ref{fig:tkioriginal}), or the $\pi + \text{proton}$ system in a pion production event, as generalized in Ref.~\cite{Lu:2019nmf} (Fig.~\ref{fig:datcompile})---the momentum of $\text{N}^\prime$ is the knocked-out nucleon momentum, and the total momentum of the pion and proton system, respectively.

The two TKI variables, transverse boosting angle~\cite{Lu:2015tcr}, $\dat$, and emulated (initial state) nucleon momentum~\cite{Furmanski:2016wqo,Lu:2019nmf}, $\pn$, are defined as:
\begin{align}
    \delta \vec{p}_\textrm{T} & = \vec{p}^{\,\lprime}_\textrm{T} + \vec{p}^{\,\nprime}_\textrm{T}, \label{eq:dpt}                                                                                                     \\
    \delta \alpha_\textrm{T}  & = \arccos\frac{ - \vec{p}^{\,\lprime}_\textrm{T} \cdot \delta \vec{p}_\textrm{T}}{\left|\vec{p}^{\,\lprime}_\textrm{T}\right| \left|\delta \vec{p}_\textrm{T}\right|}, \label{eq:dat} \\
    \delta p_\textrm{L}       & = \frac{R^2 -\delta \vec{p}_\textrm{T}^{\,2} - M^{*2} }{2R},  \label{eq:dpl}                                                                                                          \\
    \text{with:~} R           & \equiv M_\textrm{A} + p^\lprime_\textrm{L} + p^{\nprime}_\textrm{L} - E^\lprime - E^{\nprime}, \label{eq:rdef}                                                                        \\
    p_\textrm{N}              & = \sqrt{\delta \vec{p}_\textrm{T}^{\,2} +  \delta p_\textrm{L}^2} \label{eq:pn} .
\end{align}
In Eqs.~\ref{eq:dpt} and~\ref{eq:dat}, the momentum terms, $p_\text{T}^\kappa$ ($p_\text{L}^\kappa$), represent the transverse (longitudinal) component of the momentum of particle $\kappa$ in final state relative to the direction of the incoming neutrino.
$\delta \vec{p}_\textrm{T}$ is the missing momentum between initial state and final state in the transverse plane.

$\delta \vec{p}_\textrm{T}$ probes the transverse projection of the Fermi motion of the struck nucleon when the interaction proceeds via a one-body current and in the absence of FSI. Only under these conditions is its angle, $\delta \alpha_\textrm{T}$, expected to follow a uniform distribution, up to center-of-mass effects (see also Appendix~\ref{app:befor_fsi}). A deviation from uniformity in the observed $\delta \alpha_\textrm{T}$ distribution may indicate the presence of FSI or contributions from two-body currents.

In Eqs.~\ref{eq:dpl}-\ref{eq:pn}, the emulated longitudinal missing momentum, $\delta p_\textrm{L}$, is dependent on the initial nucleon mass, $M_\textrm{A}$, and the final state remnant mass, $M^*$, as well as the energies of the final-state lepton and hadron system, $E^{\lprime}$ and $E^{\nprime}$, respectively. $\pn$ is the recoil momentum of the nuclear remnant.

In the absence of FSI, $\pn$ becomes the initial struck nucleon momentum and $M^*$ can be estimated as:
\begin{equation}
    M^* = M_\textrm{A} - M_\text{n} + b,\label{eq:mstar}
\end{equation}
where $M_\text{n}$ is the neutron mass and $b$ represents the binding energy of the nucleon in the nucleus. For $^{12}\text{C}$, this binding energy is $\SI{28.7}{MeV}$. As pointed out in Ref.~\cite{MINERvA:2020anu}, the contribution of $b$ introduces only a small systematic
uncertainty (see also Appendix~\ref{app:bdep}).
When there is FSI the theoretical prediction and experimental distribution of
$\pn$ can still be compared, as the calculation remains internally consistent.

\section{\gibuu Baseline Performance}

Different kinematic regions are subject to interconnected nuclear effects, each with varying degrees of impact. Therefore, before comparing pion production data, we first examine \gibuu predictions for CC inclusive and pionless ($0\pi$) channels to establish a baseline. We anticipate that while the $\Toneb$ parameter will be constrained in this process, $\Ttpth$ will manifest primarily in pionless channels only and becomes irrelevant as current understanding indicates that 2p2h contributions do not play a role in pion production.  The newly introduced \twopibg contributes to pionless production via pion absorption through FSI, and its impact is also benchmarked.

As default in \gibuu we use the values $\Toneb = 1$, $\Ttpth = 1$, no ``Oset-type'' Delta broadening and free/vacuum FSI cross sections (as explained in Secs.~\ref{subsec:Tpar},~\ref{subsec:2p2h} and~\ref{Medium}).

\subsection{Inclusive Cross Section}

\begin{figure}[!htb]
    \includegraphics[width=\figwid]{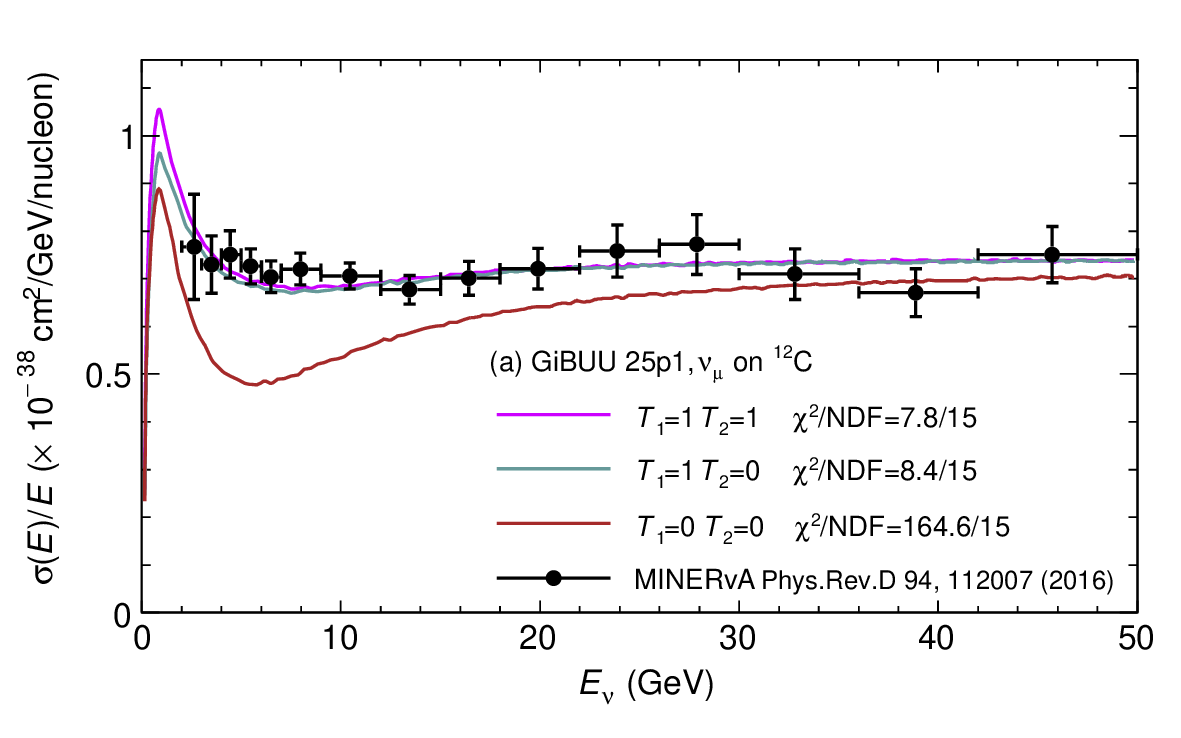}
    \includegraphics[width=\figwid]{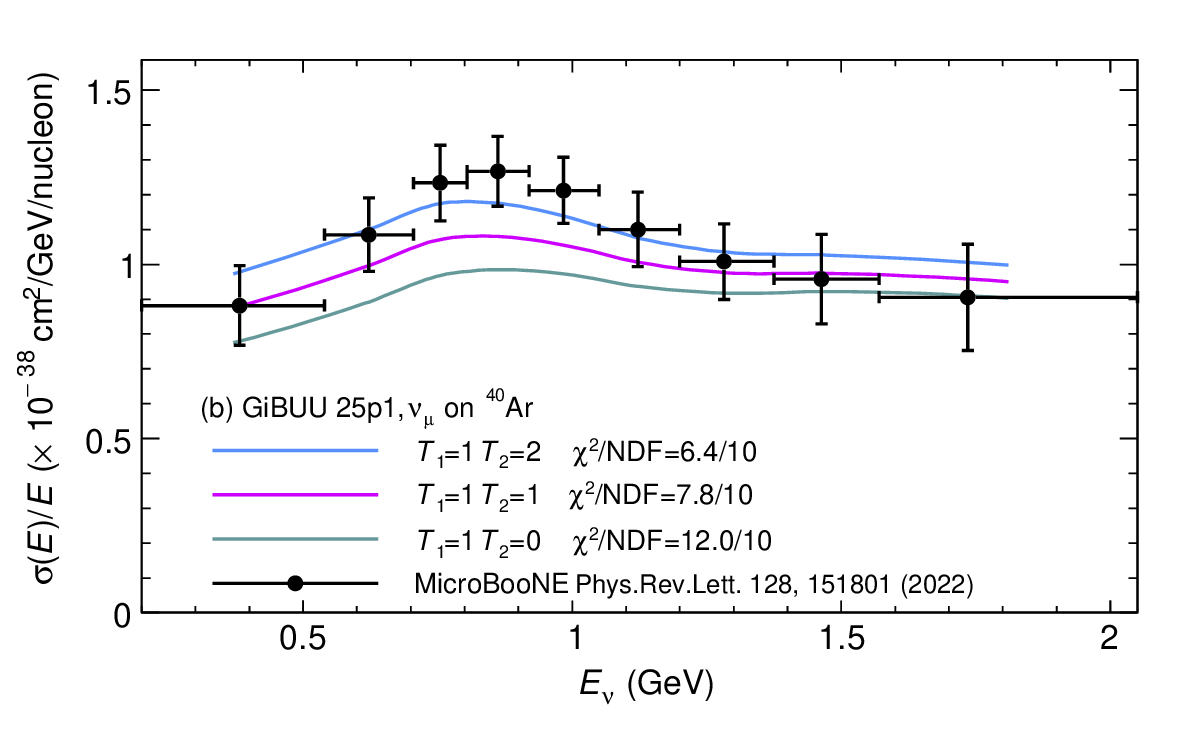}
    \caption{CC inclusive cross section measured by (a) \minerva on $^{12}\text{C}$~\cite{MINERvA:2016ing} and (b) \microboone on $^{40}\text{Ar}$~\cite{MicroBooNE:2021sfa}, compared to \gibuu predictions with different $\Tpar$ configurations.
        Note that the \microboone prediction in (b) has been forward-folded with the detector response matrix provided in the corresponding publication. A similar procedure applies to all \microboone predictions discussed in this paper. See also Appendix~\ref{app:xsec_decompose} for a channel decomposition of MINERvA predictions.
    }
    \label{fig:different_t2p2h}
\end{figure}

\minerva used the Neutrinos at the Main Injector (NuMI) Low-Energy beam---peak energy of 3 GeV with a tail extending to tens of GeV---at Fermilab and interaction targets including plastic scintillator (CH). The \minerva inclusive cross-section on  CH has been measured between 2-50 GeV~\cite{MINERvA:2016ing}.
In contrast, \microboone employed the Booster Neutrino Beam (BNB) at Fermilab, with a flux peaking at 0.8 GeV and an argon target, providing an inclusive cross section measurement up to about 2 GeV~\cite{MicroBooNE:2021sfa}.  These data are compared to \gibuu predictions in Fig.~\ref{fig:different_t2p2h}.

As mentioned in~\cref{subsec:Tpar}, in \gibuu the structure functions for neutrino interactions are obtained from the electron counterparts with a scaling factor $\left(\Tpar+1\right)$. The $\Toneb$-scaling in the one-body process primarily affects the predicted pion production at and above the second resonance region ($W > \SI{1.5}{GeV}$), where background contributions play a significant role. On the other hand, $\Ttpth$ affects only the low-energy region where the 2p2h process contributes significantly.

At energies above 2 GeV, the \minerva data in Fig.~\ref{fig:different_t2p2h}a  strongly favor $\Toneb = 1$ over $\Toneb=0$, In this energy regime ($2 < E_\nu  < 30$ GeV), pion background and SIS contributions play a major role. The cross sections obtained in this energy range with $\Toneb$ are in agreement with the typically represented isoscalar CC cross sections, as shown, for example, in Ref.~\cite{Formaggio:2012cpf}.

The 2p2h contribution to the inclusive cross section is relevant primarily at energies up to approximately 2~GeV, as shown in Fig.~\ref{fig:different_t2p2h}b. Its impact diminishes at higher energies, such as those probed by \nova and \minerva, where the effect is minimal (Fig.~\ref{fig:different_t2p2h}a). The \microboone data favor $\Ttpth = 2$,  consistent with the expected isospin value for argon~\cite{Dolan:2018sbb}. We will further probe the role of  $\Ttpth$ scaling in pionless production in the following section.

In this context it is worthwhile to note that the $A$-dependence of the 2p2h process is not well established. In most generators, and also in \gibuu per default, it is taken to be $\propto A$, appropriate for very-large-$A$ nuclei and a zero-range interaction. However, such a dependence neglects any surface effects which would lead to a steeper $A$-dependence for a short-range interaction \cite{Mosel:2016uge}. A clean distinction between the effects of the $\Ttpth$ parameter and the $A$-dependence can probably only be reached by analyzing data for very heavy nuclei.

\subsection{Pionless Production}\label{sec:cc0pi}

\minerva, T2K,  and  \microboone have conducted pioneering TKI measurements in CC0$\pi$ interactions.  \minerva~\cite{MINERvA:2018hba, MINERvA:2019ope} and \microboone~\cite{MicroBooNE:2023cmw}  used the same beam and target configurations as in their inclusive measurements, while T2K used the J-PARC beam with a narrower flux peaked at 0.6\,GeV and a hydrocarbon target~\cite{T2K:2018rnz}.
While these datasets have been used extensively to constrain nuclear effects in CC0$\pi$ interactions (see, e.g., Refs.~\cite{Yan:2024kkg, Filali:2024vpy,GENIE:2024ufm,Franco-Patino:2021yhd,Bourguille:2020bvw,Dolan:2018zye,Dolan:2018sbb}), a new combined analysis  will help benchmark \gibuu's ingredients relevant to the $W$-region for non-pion production and thus serves as a baseline for the discussions in the following sections.

The signal definition for the \minerva CC$0\pi$ measurement is:
\begin{itemize}
    \item $\nu_\mu + \text{A} \rightarrow \mu^- + \text{p} + \text{X}$, which requires a muon and at least one proton in the final state, with no pions.
    \item $\SI{1.5}{GeV/\textit{c}} < p_\mu < \SI{10}{GeV/\textit{c}}$ and $\theta_\mu < 20^\circ$.
    \item $\SI{0.45}{GeV/\textit{c}} < p_\textrm{p} < \SI{1.2}{GeV/\textit{c}}$ and $\theta_\textrm{p} < 70^\circ$.
\end{itemize}
The T2K measurement has a similar signal definition but with different phase-space cuts:
\begin{itemize}
    \item $p_\mu > \SI{0.25}{GeV/\textit{c}}$, $\cos\theta_\mu > -0.6$.
    \item $\SI{0.45}{GeV/\textit{c}} < p_\text{p} < \SI{1.0}{GeV/\textit{c}}$, $\cos\theta_\text{p} > 0.4$.
\end{itemize}
The signal definition for the \microboone CC$0\pi$ TKI measurement is:
\begin{itemize}
    \item $\nu_\mu + \text{A} \rightarrow \mu^- + \text{p} + \text{X}$, requiring a muon and one proton in the final state.
    \item Exactly one muon with $\SI{0.1}{GeV/\textit{c}} < p_\mu < \SI{1.2}{GeV/\textit{c}}$.
    \item Exactly one proton with $\SI{0.3}{GeV/\textit{c}} < p_\textrm{p} < \SI{1}{GeV/\textit{c}}$, while allowing additional protons outside this range and any number of neutrons with any momentum.
    \item No neutral pions in the final state, and all charged pions must have $p_\pi < \SI{70}{MeV/\textit{c}}$.
\end{itemize}

\begin{figure}[!htb]
    \centering
    \includegraphics[width=\figwid]{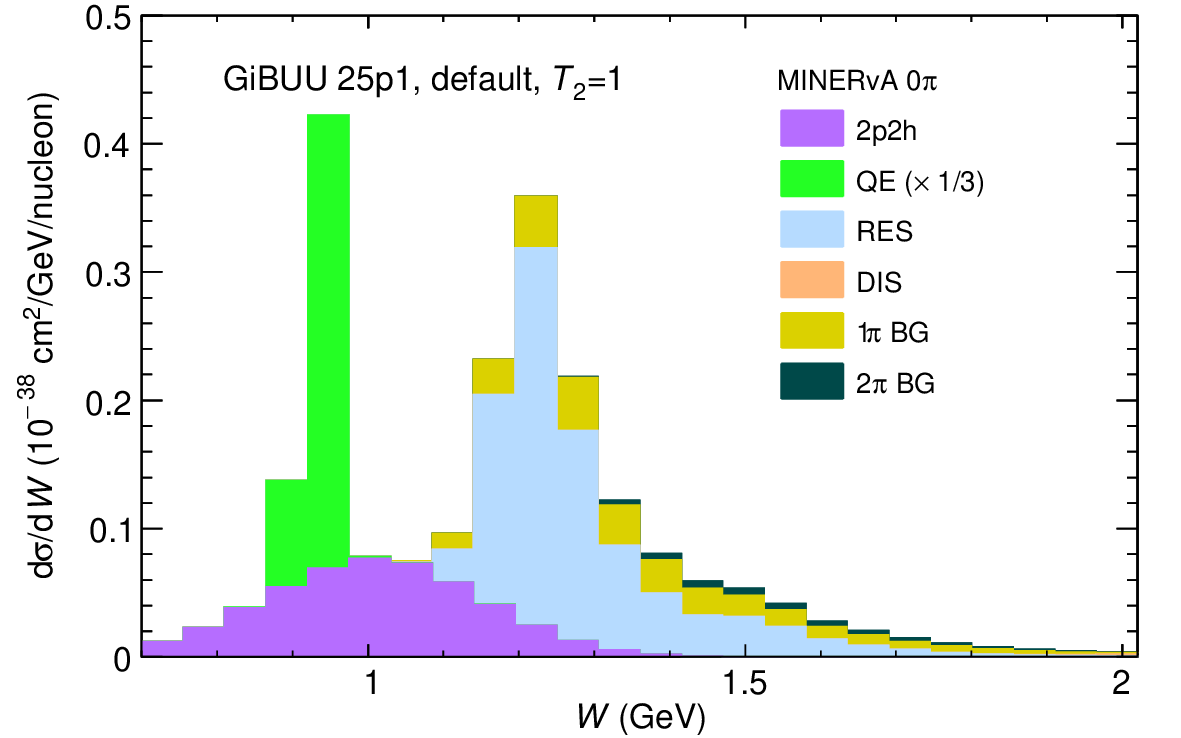}
    \caption{\gibuu  CC0$\pi$ cross section in $W$, calculated with the default configuration (plus $\Ttpth=1$) for the \minerva TKI measurement~\cite{MINERvA:2018hba}. The QE contribution is scaled by 1/3, and the DIS contribution is too small to show.}
    \label{fig:MINERvA_W_0pi}
\end{figure}

Figure~\ref{fig:MINERvA_W_0pi} shows the \gibuu cross section in $W$
\footnote{The $W$ here refers to $W_\text{true}$, defined as $W_\text{true}^2 = (p_\nu + p_\mathcal{N} - p_\ell)^2$, where $p_{\{\nu, \mathcal{N}, \ell\}}$ denotes the true four-momentum of the initial-state neutrino, initial-state nucleon, and final-state charged lepton, respectively.}
for the \minerva TKI measurement. The dominant contribution is QE and pion production, with 2p2h as the next largest contribution. Pions are produced abundantly through the $\Delta$ and higher-lying resonances at this high energy; events with pions absorbed during FSI are part of the signal. The contribution from the \twopibg compared to the one-pion background is very limited since it is less likely that two pions get reabsorbed.

\begin{figure}[!htb]
    \centering
    \includegraphics[width=\figwid]{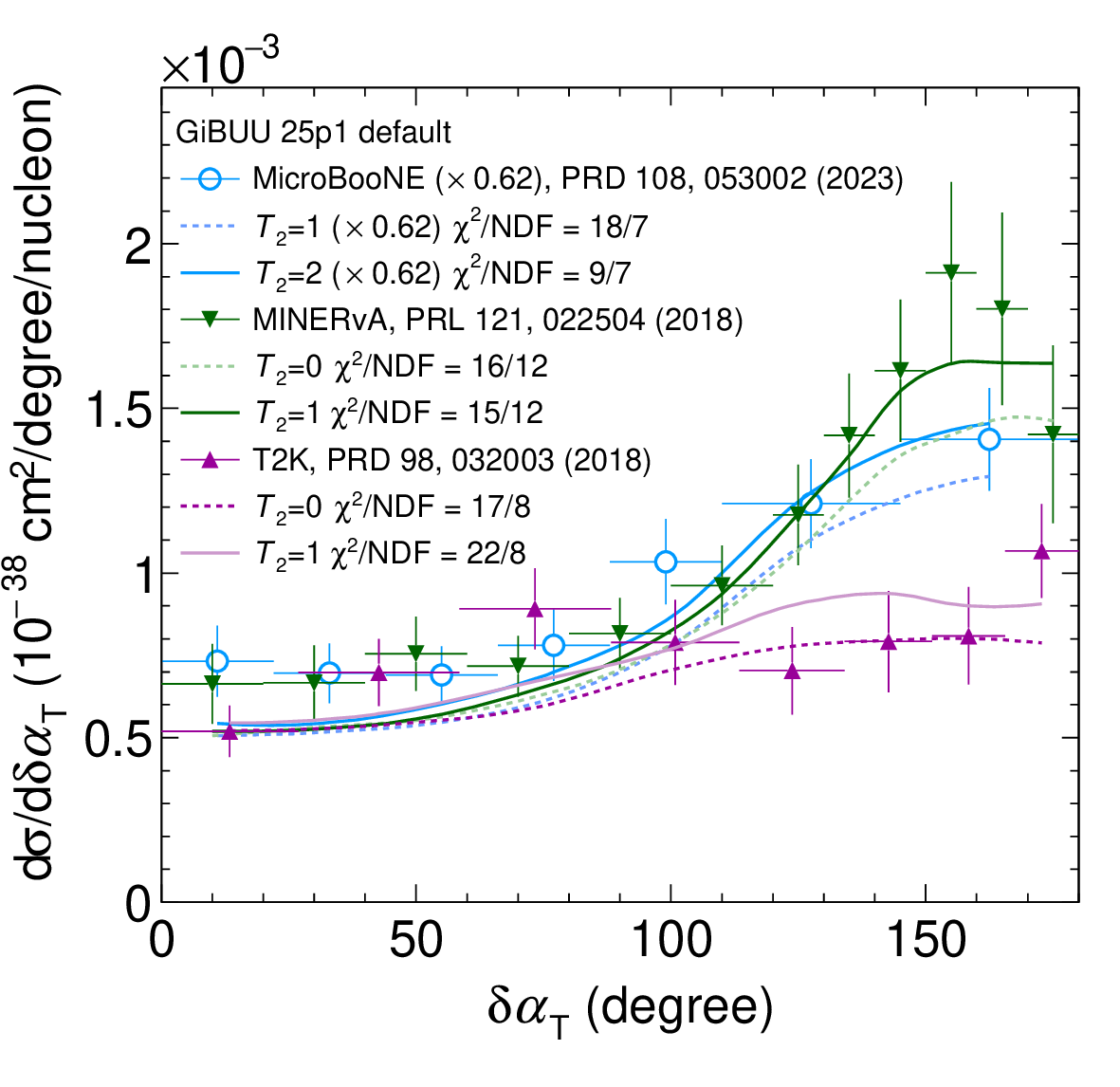}
    \caption{The CC0$\pi$ TKI cross section in $\dat$ measured by \minerva~\cite{MINERvA:2018hba,MINERvA:2019ope},  T2K~\cite{T2K:2018rnz}, and \microboone~\cite{MicroBooNE:2023cmw}, compared to the default \gibuu predictions for different $\Ttpth$ values. All data points and theory curves for \microboone are scaled by 0.62. The $\chi^2$ value for \microboone is calculated with the provided bin-by-bin errors in Ref.~\cite{MicroBooNE:2023cmw}. Comparison plots with individual channel contributions can be found in Appendix~\ref{app:minervacc0pi}. }
    \label{fig:dat_merged}
\end{figure}

The \minerva, T2K, and \microboone  CC$0\pi$ TKI measurements provide an opportunity to probe further the $\Ttpth$-scaling beyond the inclusive case in the last section. Figure~\ref{fig:dat_merged} compares the measured $\dat$ cross sections with \gibuu predictions. Overall, there is good agreement for all measurements and for all values of $\dat$. While the $\dat$ for T2K is nearly flat over the whole angular range, it shows a clear rise with angle (and thus increasing FSIs) both for \minerva and for \microboone. This different behavior directly reflects the different target sizes and energies in these experiments. Both larger target mass and higher energy lead to more final state hadron-hadron collisions thus increasing pion reabsorption.

A closer inspection shows that at small $\dat$ the cross sections are slighted underestimated, independent of the $T$ values used. There, the  cross section is dominated by pure QE interactions~\cite{Lu:2015tcr}.
In the large-$\dat$ region that is more sensitive to dissipative processes, the impact of $\Ttpth$ becomes more pronounced.

With the baseline of $\Toneb=1$, the $\chi^2$ values favor $\Ttpth=0$ for T2K. Increasing $\Ttpth$  enhances the 2p2h contribution, leading to an overshoot relative to the  data in the large-$\dat$ region.

However, the comparison shows no clear preference for \minerva, only with a mild preference towards $\Ttpth=1$. As shown in Table~\ref{tab:minervacc0pi}, the \minerva $\pn$ data of the same measurement do not distinguish the two cases.

\begin{table}[!htb]
    \centering
    \begin{tabular}{c|c|c}\toprule
        Observable                                                                   (NDF) & $\Toneb = 0$ & $\Toneb = 1$ \textbf{(default)} \\ \midrule
        \multicolumn{3}{c}{\minerva CC$0\pi$  $\Ttpth=0$}                                                                                   \\ \midrule
        $\delta \alpha_\text{{T}}$ (12)                                                    & 17           & \textbf{16}                     \\ \midrule
        $p_\text{{N}}$ (24)                                                                & 52           & \textbf{55}                     \\ \midrule
        \multicolumn{3}{c}{\minerva CC$0\pi$  $\Ttpth=1$}                                                                                   \\ \midrule
        $\delta \alpha_\text{{T}}$ (12)                                                    & 13           & \textbf{15}                     \\ \midrule
        $p_\text{{N}}$ (24)                                                                & 53           & \textbf{55}                     \\
        \bottomrule
    \end{tabular}
    \caption{$\chi^2$ between \minerva CC0$\pi$ TKI cross sections~\cite{MINERvA:2018hba,MINERvA:2019ope} and \gibuu predictions for different settings of $\Toneb$ and $\Ttpth$. The default setting is in bold font and the corresponding plots can be found in Appendix~\ref{app:minervacc0pi}. }\label{tab:minervacc0pi}
\end{table}

For a carbon target, we expect $\Ttpth=0$ based on the naive isospin scaling, following the same scaling rule applied to the argon case in the previous section on the inclusive measurements. However, comparison with the CC$0\pi$ data presented here indicates that this $\Ttpth$-scaling does not hold consistently across different energy regimes.

For \microboone, we expect  $\Ttpth=2$ as discussed in Sec.~\ref{subsec:Tpar}. The observed data in Fig.~\ref{fig:dat_merged} confirm this expectation, as they clearly prefer $\Ttpth=2$ to $\Ttpth=1$.

In the following discussion, we use $\Ttpth=1$ as the nominal value. It should be emphasize that  this parameter has nearly zero effect on pion production predictions, due to the absence of 2p2h contributions in that channel; pions could be produced only through final state interactions.

In Table~\ref{tab:minervacc0pi}, we also compare the \minerva data to the \gibuu calculations using different $\Toneb$ settings. Overall,  the results do not indicate a strong preference.

\section{Neutrino Pion Production}\label{sec:pipro}

As discussed in Sec.~\ref{subsec:pion}, neutrino-induced pion production on nuclei in the resonance region is governed by different mechanisms, including resonance production and non-resonant background processes. In this section, we present the model-data comparison for CC pion production in \minerva and \microboone. For neutral-current (NC) processes on MicroBooNE, a cross section analysis in \gibuu has been performed in Ref.~\cite{bogart2024inmedium}.

\subsection{CC$\pi^0$ Production on Carbon}\label{subsec:ccpi0}

\minerva measured the CC$\pi^0$ TKI cross section~\cite{MINERvA:2020anu} with the same beam and target as the above inclusive and pionless measurement.
The signal definition is:
\begin{itemize}
    \item $\nu_\mu + \text{A} \rightarrow \mu^- + \text{p} + \pi^0 + \text{X}$, which requires one muon and at least one proton and at least one $\pi^0$ in the final state, with no charged pions.
    \item $\SI{1.5}{GeV/\textit{c}} < p_\mu < \SI{20}{GeV/\textit{c}}$ and $\theta_\mu < 25^\circ$.
    \item $p_\mu > \SI{0.45}{GeV/\textit{c}}$.
\end{itemize}

\begin{figure}[!htb]
    \centering
    \includegraphics[width=\figwid]{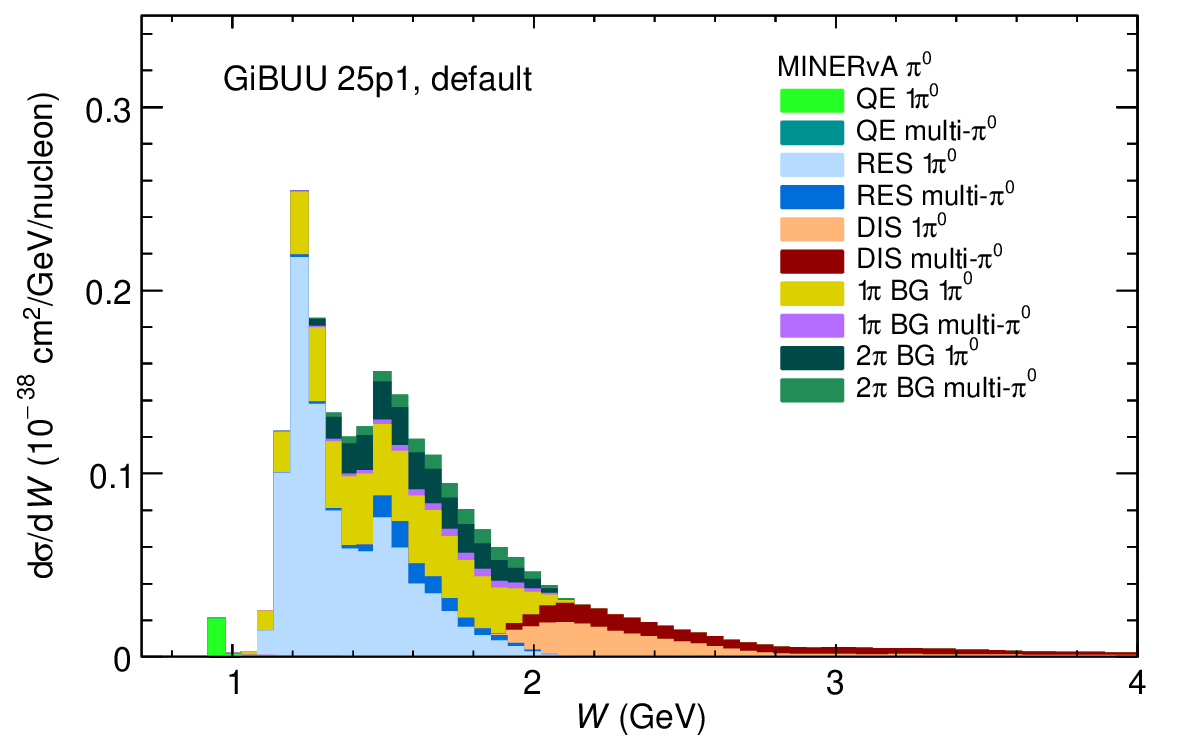}
    \caption{\gibuu cross section in $W$ for the \minerva CC$\pi^0$ measurement~\cite{MINERvA:2020anu}.}
    \label{fig:MINERvA_W_pi0}
\end{figure}

The default \gibuu prediction for ${\rm d}\sigma/{\rm d} W$  for the \minerva CC$\pi^0$ measurement is shown in \figref{fig:MINERvA_W_pi0}. It  clearly illustrates the contributions from different interaction channels across various kinematic regions. Compared to the CC0$\pi$ case in Fig.~\ref{fig:MINERvA_W_0pi}, the 2p2h contribution is absent here.
The total contribution includes QE interactions ($W$ at the nucleon mass) via $\pi$ production during FSI, $\Delta$ production around $W = \SI{1.2}{GeV}$, the second resonance region at $W = \SI{1.5}{GeV}$, and DIS contributions at higher~$W$. In the resonance region, contributions from both $1\pi$ and \twopibg are significant. Between the resonance and DIS regions, a smooth transition is observed.

\begin{figure}[!htb]
    \centering
    \includegraphics[width=\figwid]{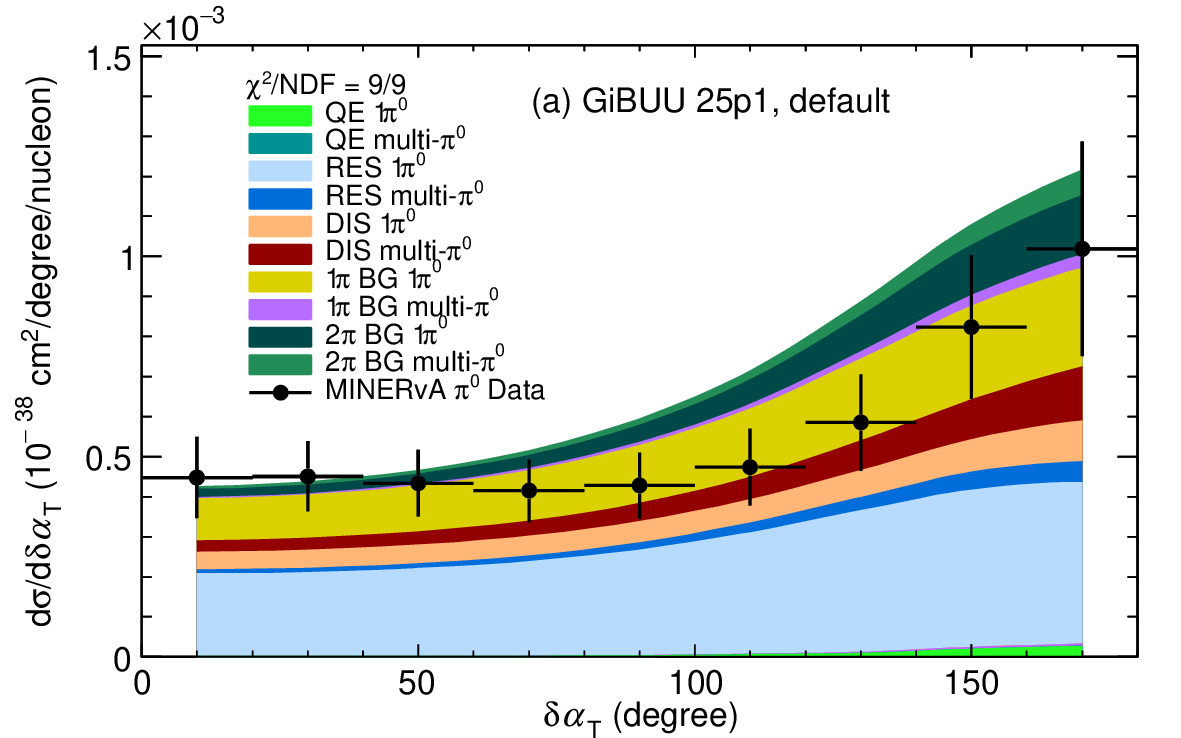}
    \includegraphics[width=\figwid]{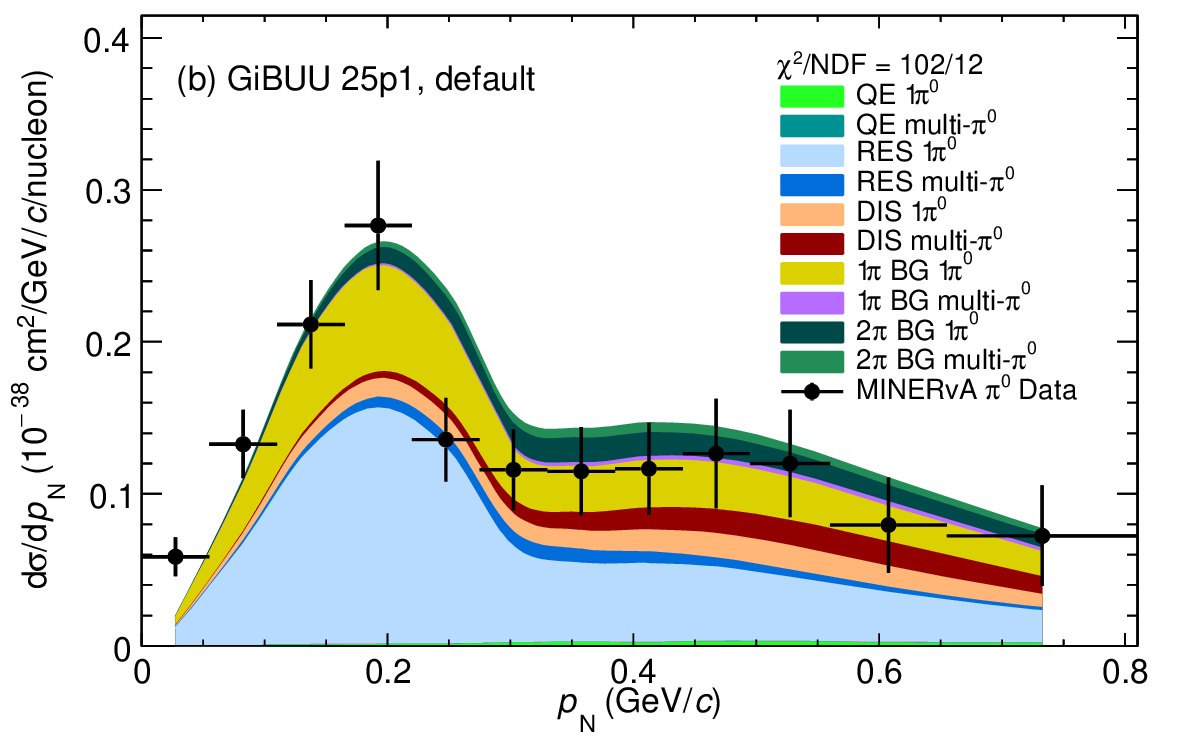}
    \caption{\minerva  CC$\pi^0$ TKI cross section~\cite{MINERvA:2020anu} in (a) $\dat$ and (b) $\pn$ compared to \gibuu predictions with $\Ttpth=1$.  }    %
    \label{fig:pi0T0}
\end{figure}

The momenta of the selected $\pi^0$ and proton are combined and  correlated with that of the muon to construct the TKI observables~\cite{Lu:2019nmf,MINERvA:2020anu}.
The measured cross sections and the default \gibuu predictions are compared in Fig.~\ref{fig:pi0T0}. For $\dat$, the model provides a good description in the non-dissipative region ($\dat \to 0$), but increasingly overshoots the data at larger $\dat$ values.
For $\pn$, the location of the Fermi motion peak is well reproduced; however, the shape of the peak is not, with the falling edge at the Fermi surface (0.2-0.3~GeV/$c$) exhibiting excess strength.

\subsection{CC$\pi^0$ Production on Argon}\label{subsec:ccpi0_ar_microboone}

\begin{figure}[!htb]
    \centering
    \includegraphics[width=\figwid]{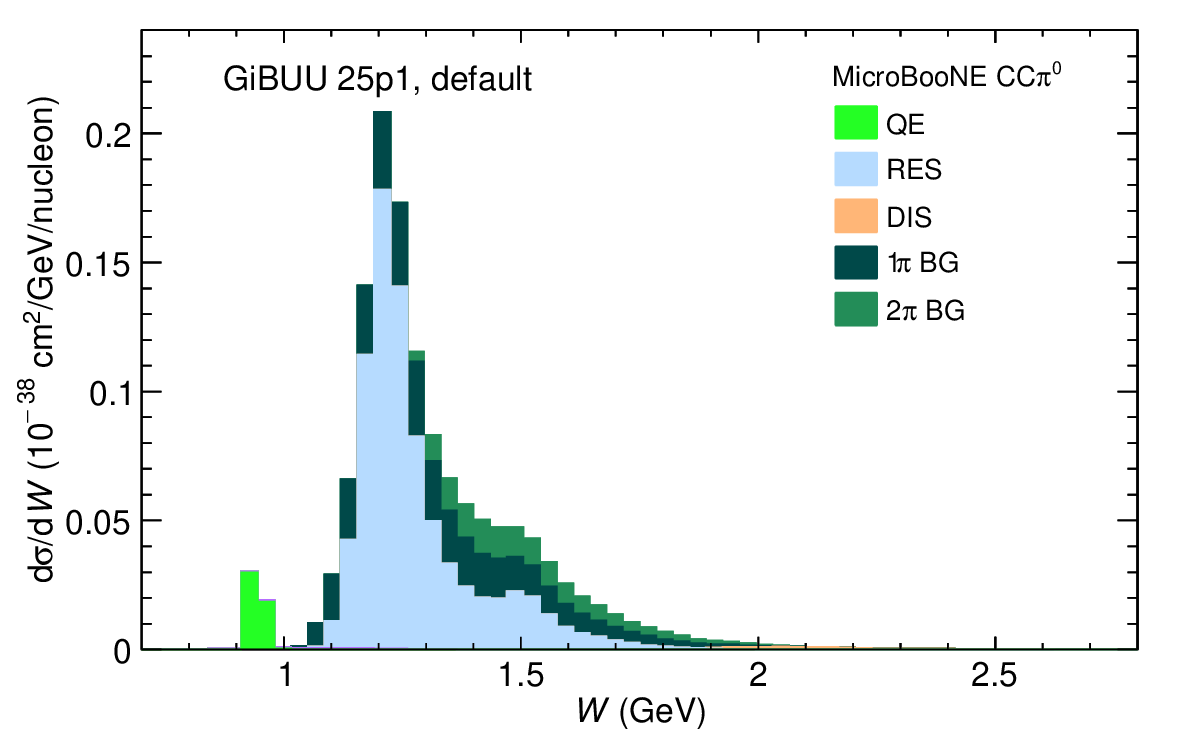}
    \caption{\gibuu cross section in $W$ for the \microboone CC$\pi^0$ measurement~\cite{microboone:pi0}.}   \label{fig:microboone_W}
\end{figure}

\microboone measured the CC$\pi^0$ cross sections in the $\pi^0$ kinematics~\cite{microboone:pi0} with the same beam and target as the inclusive measurement. It requires the final state to contain exactly one muon, one $\pi^0$, any number of nucleons, and no other leptons or hadrons. Because of the low beam energy, $\pi^0$ production at \microboone occurs predominantly through $\Delta$ resonance, with minor contributions from the second resonance region and DIS, as shown in Fig.~\ref{fig:microboone_W}. Nevertheless, the contributions from the $1\pi$BG or \twopibg are significant; the latter enters the sample because of pion absorption during FSI---one pion is absorbed and the surviving one is a $\pi^0$.

\begin{figure}[!htb]
    \centering
    \includegraphics[width=\figwid]{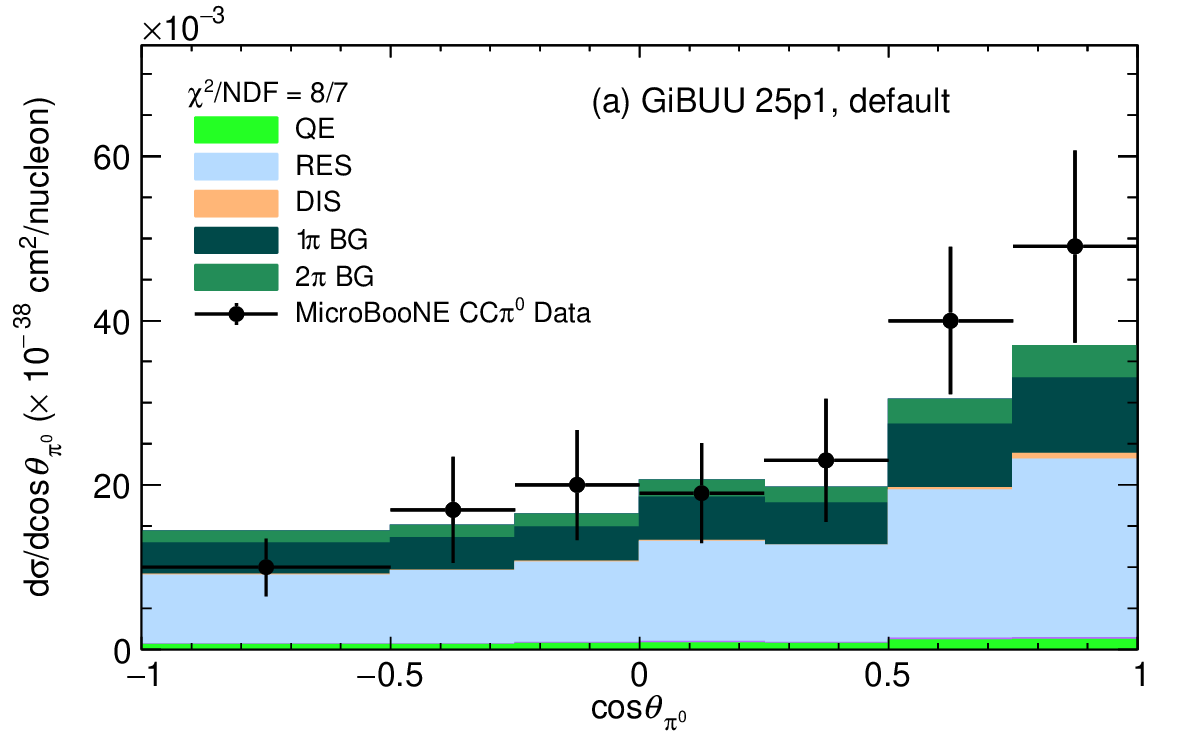}
    \includegraphics[width=\figwid]{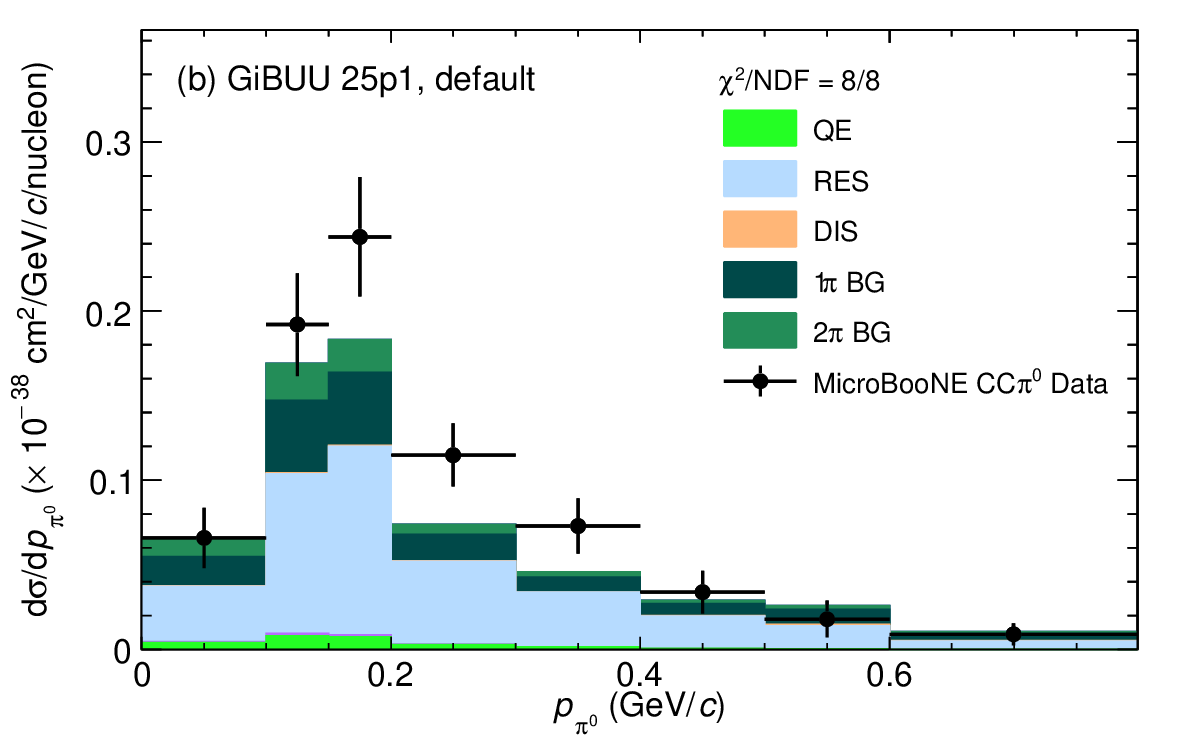}
    \caption{MicroBooNE CC$\pi^0$ cross sections~\cite{microboone:pi0} in (a) $\ubang$ and (b) $\ubmom$, the contribution from different interaction channels are decomposed.  }
    \label{microbooneplot:pi0}
\end{figure}

For the $\pi^0$ scattering angle in Fig.~\ref{microbooneplot:pi0}a, we can see a underestimation of the \gibuu prediction in the forward region. For the momentum distribution in Fig.~\ref{microbooneplot:pi0}b, \gibuu is below the data between 100-200~MeV/$c$. It is interesting to note that the very same discrepancies were also observed in a \gibuu analysis of MicroBooNE NC pion production data \cite{bogart2024inmedium}. We will come back to these discrepancies when we discuss the influence of possible in-medium changes on these results in Sec.~\ref{sec:systematics}.

Unlike the inclusive cross sections, it is not possible to simultaneously describe both \minerva~\cite{MINERvA:2020anu} (Fig.~\ref{fig:pi0T0} above) and \microboone~\cite{microboone:pi0} CC$\pi^0$ production with the current \gibuu model. %
Notably, the recent hybrid model in \nuwro~\cite{Yan:2024kkg} also fails to reproduce the \microboone measurement, with a significant underestimate in the similar kinematic regions.

\section{\gibuu in-medium effects}\label{sec:systematics}

In the previous sections, the \gibuu default predictions do not include the Oset broadening of the $\Delta$-resonance width, assume free  FSI cross section, and set $\Toneb=1$ while $\Ttpth$ is treated as an effective parameter. The treatment of both the  $\Delta$-resonance width and  FSI cross section introduce systematic uncertainties.
In this section, we investigate the impact of these in-medium effects. The resulting $\chi^2$ values are summarized in Tables~\ref{tab:minervasyschi2}  for the aforementioned \minerva CC0$\pi$ TKI, \minerva CC$\pi^0$ TKI, and \microboone CC$\pi^0$ measurements.

\begin{table}[!htb]
    \centering
    \begin{tabular}{c|c|c|c|c}\toprule
        \makecell{Observable                                            \\ (NDF)} & \makecell{\textbf{w/o Oset} \\ \textbf{free } \\\textbf{(default)}} &  \makecell{w/ Oset \\ free } & \makecell{w/ Oset \\ in-med } & \makecell{w/o Oset \\ in-med } \\ \midrule
        \multicolumn{5}{c}{\minerva CC$0\pi$   $\Ttpth=0$}              \\ \midrule
        $\delta \alpha_\text{{T}}$ (12) & \textbf{16}  & 16 & 14  & 16  \\ \midrule
        $p_\text{{N}}$ (24)             & \textbf{55}  & 82 & 69  & 78  \\ \midrule
        \multicolumn{5}{c}{\minerva CC$0\pi$   $\Ttpth=1$}              \\ \midrule
        $\delta \alpha_\text{{T}}$ (12) & \textbf{15}  & 18 & 15  & 14  \\ \midrule
        $p_\text{{N}}$ (24)             & \textbf{55}  & 87 & 73  & 69  \\ \midrule

        \multicolumn{5}{c}{\minerva CC$\pi^0$ $\Ttpth=1$}               \\ \midrule
        $\delta \alpha_\text{{T}}$ (9)  & \textbf{9}   & 6  & 9   & 15  \\ \midrule
        $p_\text{{N}}$ (12)             & \textbf{102} & 81 & 124 & 153 \\ \midrule

        \multicolumn{5}{c}{\microboone CC$\pi^0$ $\Ttpth=1$}            \\ \midrule
        $\ubang$ (7)                    & \textbf{8}   & 7  & 6   & 20  \\ \midrule
        $\ubmom$ (8)                    & \textbf{8}   & 22 & 13  & 7   \\

        \bottomrule
    \end{tabular}
    \caption{$\chi^2$ between experimental data and \gibuu predictions for different medium configurations. From upper to lower rows:
        \minerva CC0$\pi$ TKI cross sections~\cite{MINERvA:2018hba,MINERvA:2019ope} with different $\Ttpth$,  \minerva CC$\pi^0$ cross sections~\cite{MINERvA:2020anu}, and \microboone CC$\pi^0$ cross sections~\cite{microboone:pi0}.
        The default setting is in bold font.}
    \label{tab:minervasyschi2}
\end{table}

\begin{figure}[!b]
    \centering
    \includegraphics[width=\figwid]{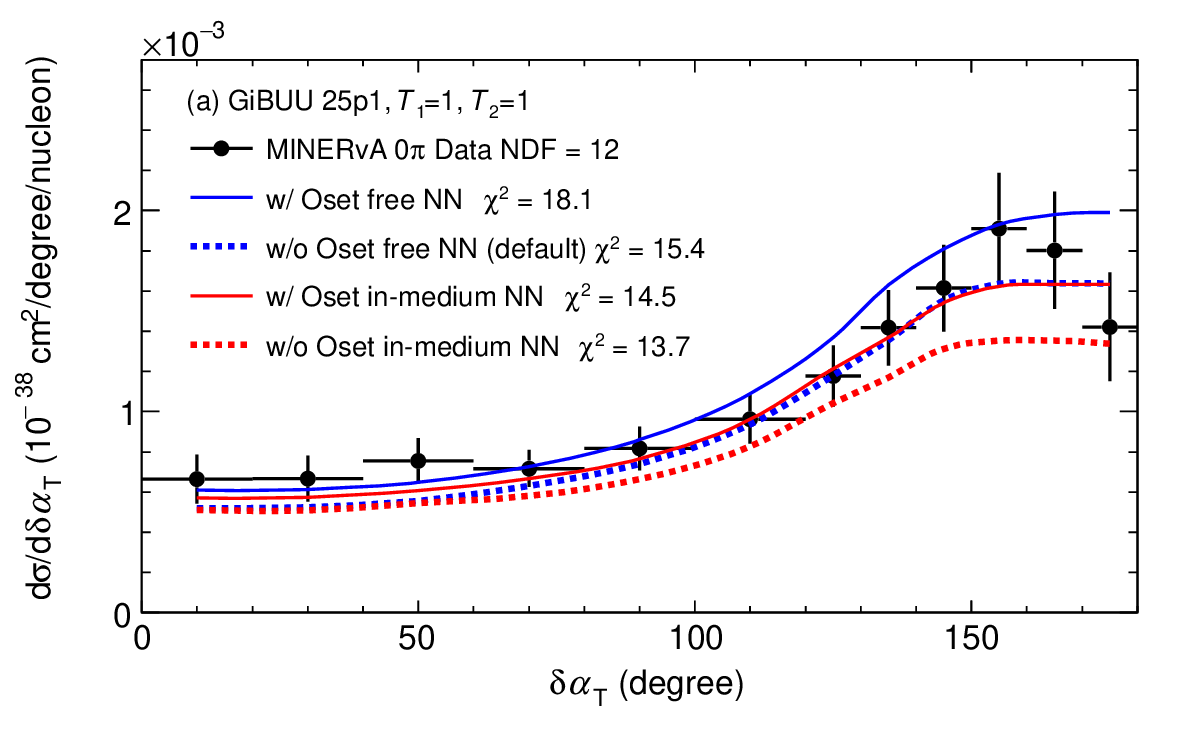}
    \includegraphics[width=\figwid]{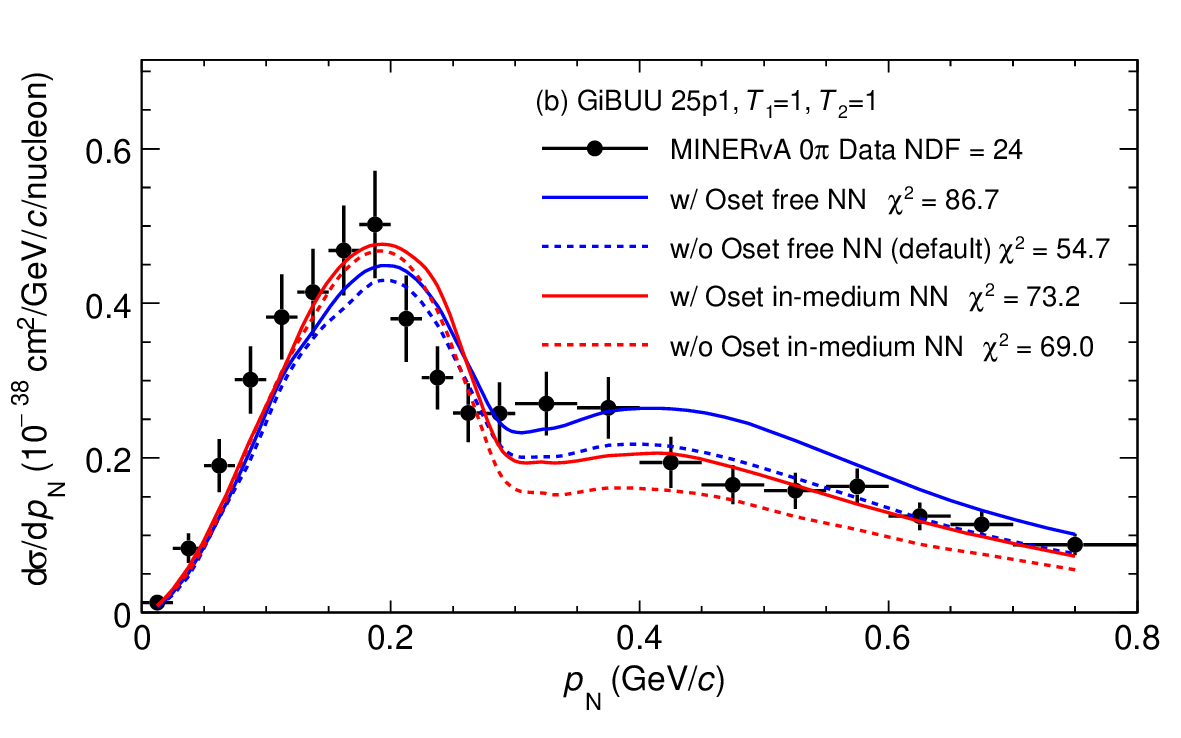}
    \caption{\minerva CC0$\pi$ TKI cross sections~\cite{MINERvA:2018hba} compared to \gibuu predictions with different medium configurations.}\label{fig:inmedium_0pi}
\end{figure}

As shown in Table~\ref{tab:minervasyschi2}, the \minerva CC0$\pi$ TKI measurement favors the default setting---without Oset broadening and with the free  FSI cross section---regardless of $\Ttpth$; the data-GiBUU comparison for $\Ttpth=1$ is shown in Fig.~\ref{fig:inmedium_0pi}. There are three interesting observations for this CC0$\pi$ sample. First, the impact of in-medium variations is much more pronounced in the dissipative regions, i.e. $\dat\to 180^\circ$ and $\pn>0.3~\textrm{GeV}/c$. Second, the Oset broadening increases the cross section for 0$\pi$ events because it lowers the pion production cross section. On the other hand, the in-medium  cross section reduces pion absorption because the in-medium change involves a lowering of pion reabsorption. Thus these two effects approximately cancel each other.

\begin{figure}[!b]
    \centering
    \includegraphics[width=\figwid]{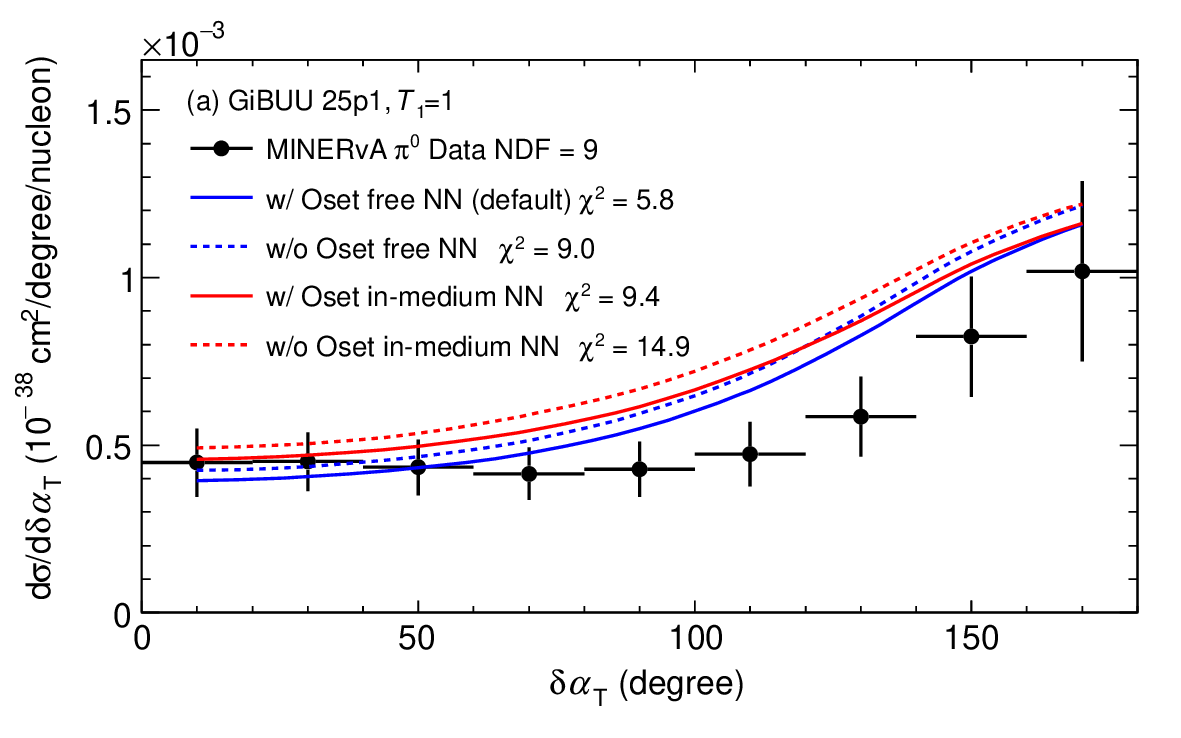}
    \includegraphics[width=\figwid]{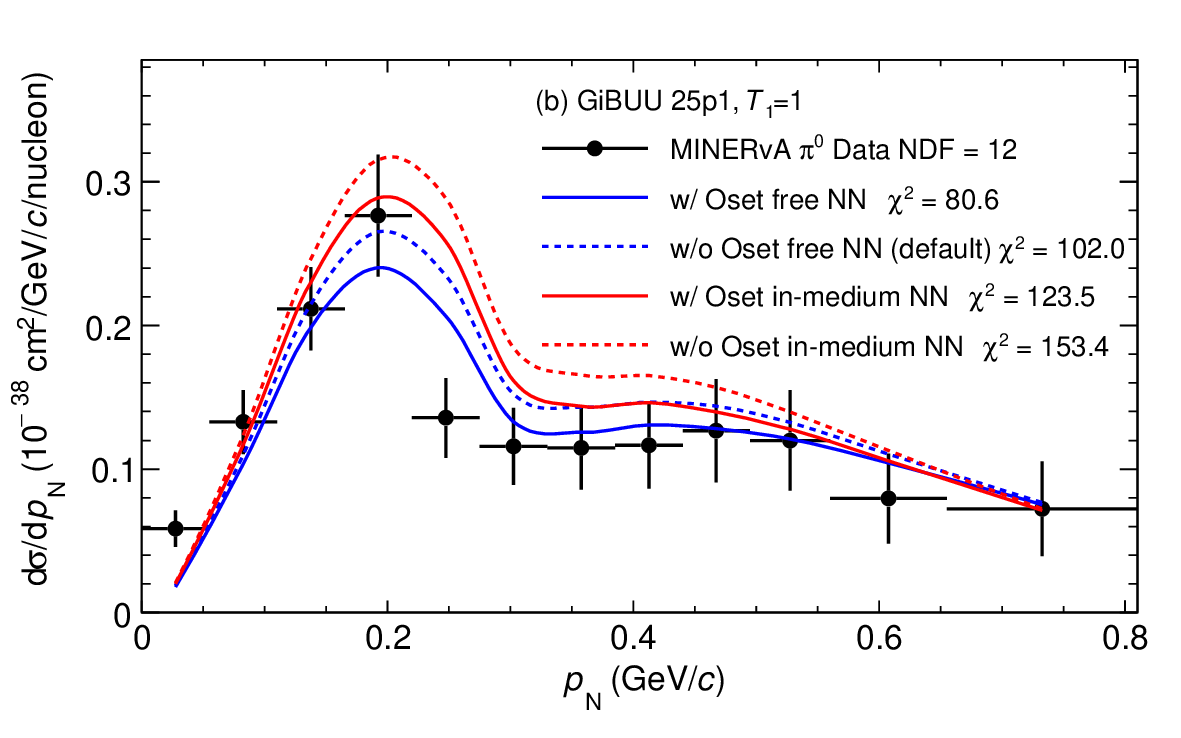}
    \caption{\minerva CC$\pi^0$ TKI cross sections~\cite{MINERvA:2020anu} compared to \gibuu predictions with different medium configurations: ``Oset'' refers to Oset broadening.  } \label{fig:minerva_alt}
\end{figure}

When applied to the \minerva CC$\pi^0$ measurement, the impact of these variations is mild, and not in the dissipative regions as in the CC0$\pi$ case above. The integrated cross section is enhanced by $-6\%$ to $+8\%$, with the minimum corresponding to the configuration with Oset broadening and free  cross section, and the maximum to the configuration without broadening and with in-medium  cross section---an opposite ordering as the  CC0$\pi$ case.
As shown in \figref{fig:minerva_alt}, both $\dat$ and $\pn$ show a stronger preference for the minimum in-medium enhancement with Oset broadening  and free  cross section over the default, as this yields the lowest cross section, thereby reducing the degree to which the prediction overshoots the data.

\begin{figure}[!htb]
    \centering
    \includegraphics[width=\figwid]{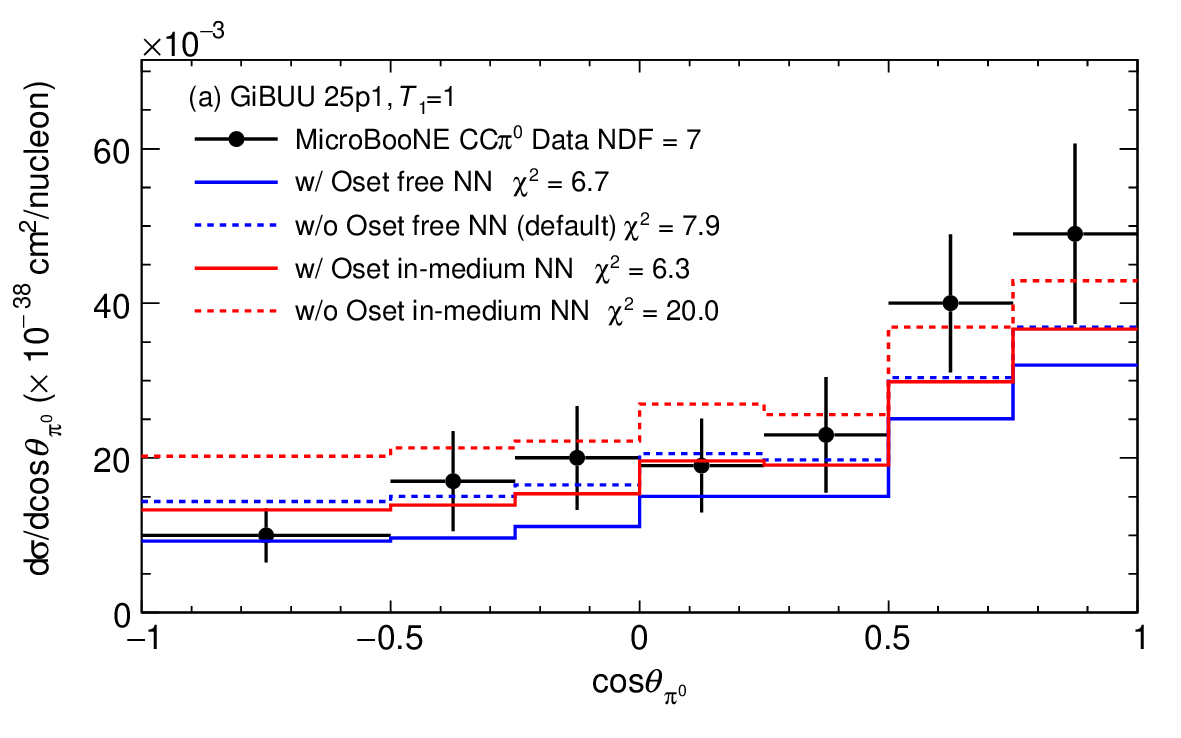}
    \includegraphics[width=\figwid]{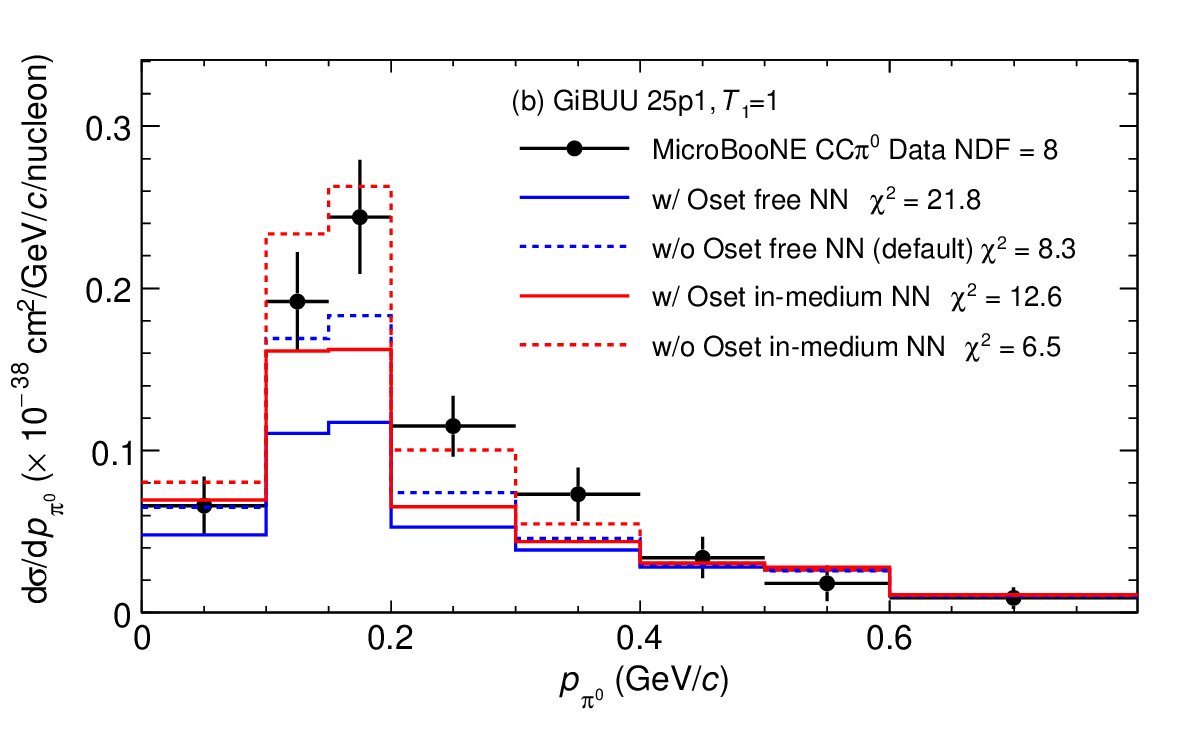}
    \caption{\microboone CC$\pi^0$ cross sections in $\ubang$ and $\ubmom$~\cite{microboone:pi0} compared to \gibuu predictions with different medium configurations.   } \label{fig:microboone_alt}
\end{figure}

For the \microboone CC$\pi^0$ measurement,  in-medium effects significantly alter the cross section, as shown in \figref{fig:microboone_alt}, due to argon's higher nuclear density relative to carbon. The integrated cross section varies by $-25\%$ to $+29\%$, with the  configurations giving the extrema being the same as those for the \minerva CC$\pi^0$ TKI measurement discussed above.
At the peak of the $\ubmom$ distribution, the variation reaches up to approximately $\pm 40\%$.  However, the ordering of configurations from minimum to maximum differs:
for \microboone, the next-to-minimum result comes from the configuration with Oset broadening and in-medium  cross sections, followed by the default, whereas for \minerva, the default yields the next-to-minimum.

We therefore conclude that, in CC$\pi^0$ production, the effects of Oset broadening and in-medium  cross sections on the production are opposite in sign and comparable at leading order, with nuclear dependence introducing a difference at subleading order. This overall behavior is consistent with the CC0$\pi$ case, except for the sign of the impact because the Oset broadening enhances pion reabsorption, whereas the in-medium  cross section reduces it.

As discussed in Sec.~\ref{subsec:ccpi0_ar_microboone}, the default \gibuu predictions underestimate the cross sections in the forward region, $\ubang > 0.5$, and in the peak region of $\ubmom$ (100–200~MeV/$c$). Figure~\ref{fig:microboone_alt} shows that the maximum in-medium enhancement recovers the deficit but leads to an overshoot in the backward region of $\ubang$. Thus, the maximum in-medium enhancement is favored by the $\ubmom$ distribution but strongly disfavored by the $\ubang$ distribution.

In a recent study comparing \gibuu to the \microboone NC $\pi^0$ data~\cite{MicroBooNE:2024pdj, bogart2024inmedium}, a similar observation was made: the deficits in the $\ubmom$ and $\ubang$ distributions can be mitigated by the same in-medium effects---namely, disabling Oset broadening while lowering pion reabsorption by enabling the in-medium  cross section . However, since \gibuu does not model coherent pion production, which contributes to the NC$\pi^0$ signal, the authors of Ref.~\cite{bogart2024inmedium} concluded that the apparent improvement may be compensating for the missing coherent component.

In the present case of CC$\pi^0$ production, where coherent pion production is absent, this ambiguity does not arise. Nonetheless, an inconsistent picture now emerges: the \minerva CC$\pi^0$ measurement on carbon prefers minimal in-medium enhancement; the \microboone CC$\pi^0$ measurement on argon requires maximal in-medium enhancement, albeit with unresolved issues in the angular distribution; and the \microboone NC$\pi^0$ measurement on argon appears to require an in-medium enhancement that is fine-tuned to be the maximum minus the coherent production cross section. This inconsistent picture could be clarified once pion TKI analyses for \microboone (or SBND) become available.

Summarizing this section we note again  that there is an interplay between the Oset-type broadening of the $\Delta$ resonance and in-medium effects, affecting the pion absorption cross sections. A final decision on whether both or only one process should always be turned on (or off) is difficult. The $\Delta$-broadening has been checked only for rather low energy electron and photon reactions and the $\Delta$-suppression of Eq. (\ref{Delta-supp}) has been taken over from nucleus-nucleus and p-nucleus reactions on pion production; in these reaction types problems still exist in the experimental results on pion production \cite{Kummer:2023hvl}. It would, therefore, be interesting to see an appllication of results of a consistent calculation of both the $\Delta$ in-medium broadening and of the $\Delta$ absorption cross section as performed in Ref.~\cite{Larionov:2003av} for the description of heavy-ion reactions.

\section{Summary}\label{sec:summary}

In this work, we use the most up-to-date \gibuu model (version \gibver) to study neutrino-induced pion production measured by \minerva~\cite{MINERvA:2020anu} and \microboone~\cite{microboone:pi0}. We begin by evaluating the baseline performance of \gibuu against inclusive measurements from \minerva~\cite{MINERvA:2016ing} and \microboone~\cite{MicroBooNE:2021sfa}, as well as pionless TKI data from \minerva~\cite{MINERvA:2021csy}, T2K~\cite{T2K:2018rnz},  and~\microboone~\cite{MicroBooNE:2023cmw}. In our analysis, we relate our findings to a \nuwro analysis of \microboone's CC$\pi^0$~\cite{Yan:2024kkg,microboone:pi0} data and a \gibuu analysis of \microboone's NC$\pi^0$ production~\cite{bogart2024inmedium,MicroBooNE:2024pdj}.

An important aspect of the \gibuu model is that it scales background contributions to electron scattering cross sections to neutrino cross sections. This scaling has recently been decoupled for pion production via the one-body current and for 2p2h contributions, using separate scaling factors $\Toneb$ and $\Ttpth$, respectively. While inclusive measurements favor $\Toneb = 1$, we found that $\Ttpth = 2$ better describes the \microboone inclusive measurement, whereas $\Ttpth = 1$ and 0 are preferred by the \minerva and T2K CC0$\pi$ TKI measurements, respectively. Since the 2p2h contributions have only a very small effect on pion production, the value of $\Ttpth$ does not affect our present study.

We observe a tension in the data: the \minerva CC0$\pi$ TKI measurement does not seem to require the $2\pi$ background contribution, as the prediction already tends to overestimate the data without it; in contrast, the \microboone CC$\pi^0$ data appear to lack sufficient strength in this component, with the model exhibiting a significant deficit---a deficit also seen in the \nuwro analysis~\cite{Yan:2024kkg}.

The relevant in-medium systematic effects of \gibuu include Oset broadening and  FSI cross sections. Deviation from the default vacuum case---no broadening and free  cross section---leads to the in-medium enhancement. While the default case is established in the \minerva CC0$\pi$ TKI measurement, the situation for pion production remains unsettled.

In both the \minerva and \microboone CC$\pi^0$ data, Oset broadening reduces pion production while in-medium cross sections increase it by weakening pion reabsorption, with comparable first-order magnitudes. The combinations produce the minimum enhancement (effective reduction) with Oset broadening and free  cross section, and the maximum enhancement without Oset broadening and with in-medium  cross section.
This mirrors the \twopibg observation: \minerva CC data seem to require minimal enhancement while \microboone (both CC and NC~\cite{bogart2024inmedium}) prefers maximal. The angular distributions and relation to coherent pion production remain unresolved.
The tension might also arise from a possible energy dependence of these effects, since the underlying $W$ distributions differ significantly between the two experiments, as shown in \cref{fig:MINERvA_W_pi0} and \cref{fig:microboone_W}.

In this paper we have tried to extract fundamental aspects of neutrino interactions from different experiments involving different targets and different neutrino energy regimes, concentrating on the pion production channel. The present study thus extends the work of Ref.~\cite{Mosel:2023zek} which was restricted to inclusive cross sections. However, the apparent inconsistencies between datasets present a challenge. One possible approach within the TKI framework identifies specific nuclear effects manifesting in distinct regions. For example, the non-dispersive region ($\dat \to 0$) exhibits universal simplicity, with contributions devoid of FSI and unaccounted-for particles. Consistency among different models, event topology, and datasets could first be achieved in this region as a calibration step, as shown in Figs.~\ref{fig:datcompile} and~\ref{fig:dat_merged}. With new analyses and data emerging---such as those from SBND~\cite{SBND:2020scp,SBND:2024vgn} and T2K~\cite{T2K:2019bbb}---such systematic approaches may provide conclusive indications for model improvement.

\begin{acknowledgments}
    We thank Mariam~Rifai for discussions.
    Q.Y. and Y.Z. are supported by National Natural Science Foundation of China (NSFC) under contract 12221005.
\end{acknowledgments}

\bibliography{bibliography}%

\clearpage

\appendix

\renewcommand\thefigure{\thesection.\arabic{figure}}

\renewcommand\thetable{\thesection.\arabic{table}}

\section{\minerva TKI prediction before FSI}\label{app:befor_fsi}
\begin{figure}[!htb]
    \includegraphics[width=0.4\textwidth]{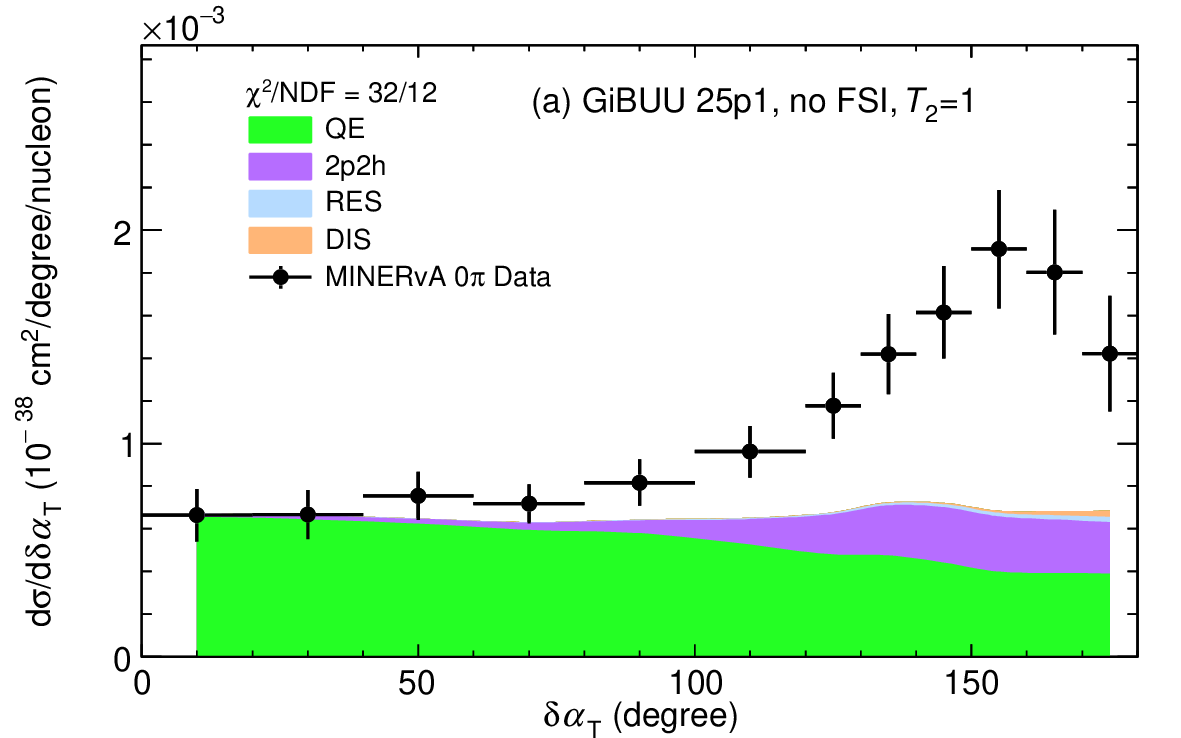}
    \includegraphics[width=0.4\textwidth]{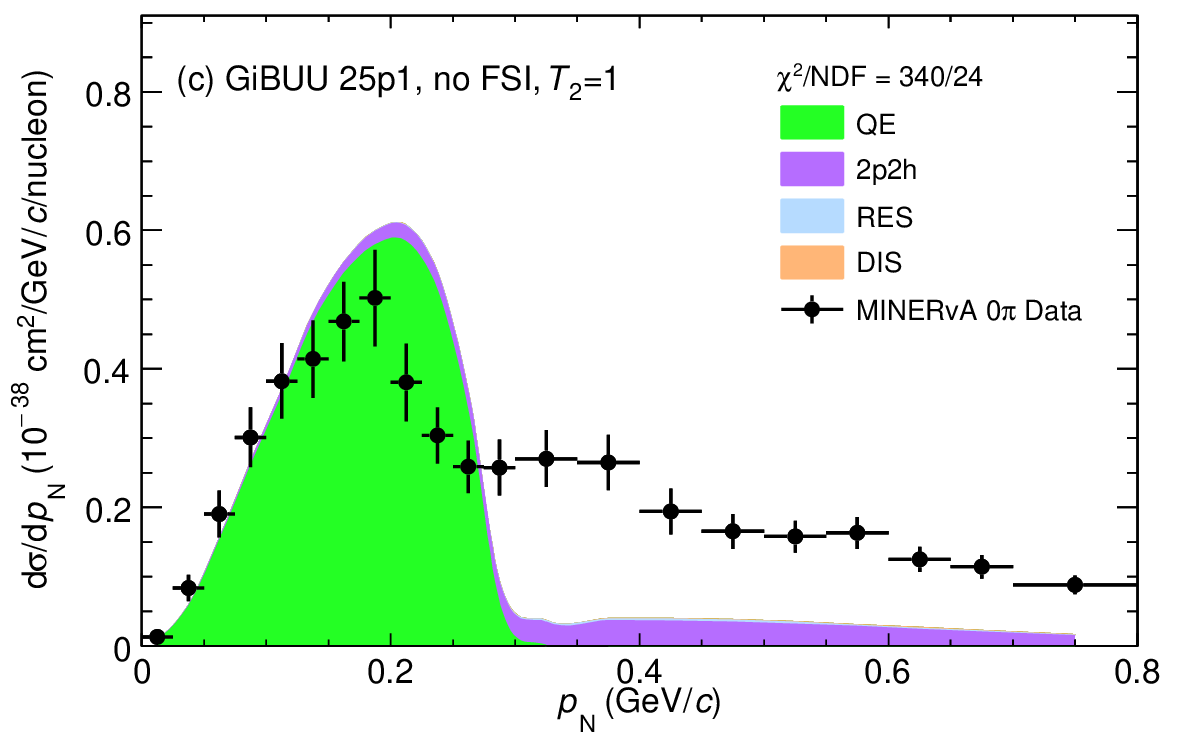}
    \caption{\gibuu CC0$\pi$ TKI cross sections in (a) $\dat$ and (b) $\pn$~\cite{MINERvA:2018hba,MINERvA:2019ope} calculated before FSI, the contribution from different interaction channels are decomposed.  } \label{fig:nofsi_0pi}
\end{figure}

\begin{figure}[!htb]
    \includegraphics[width=0.4\textwidth]{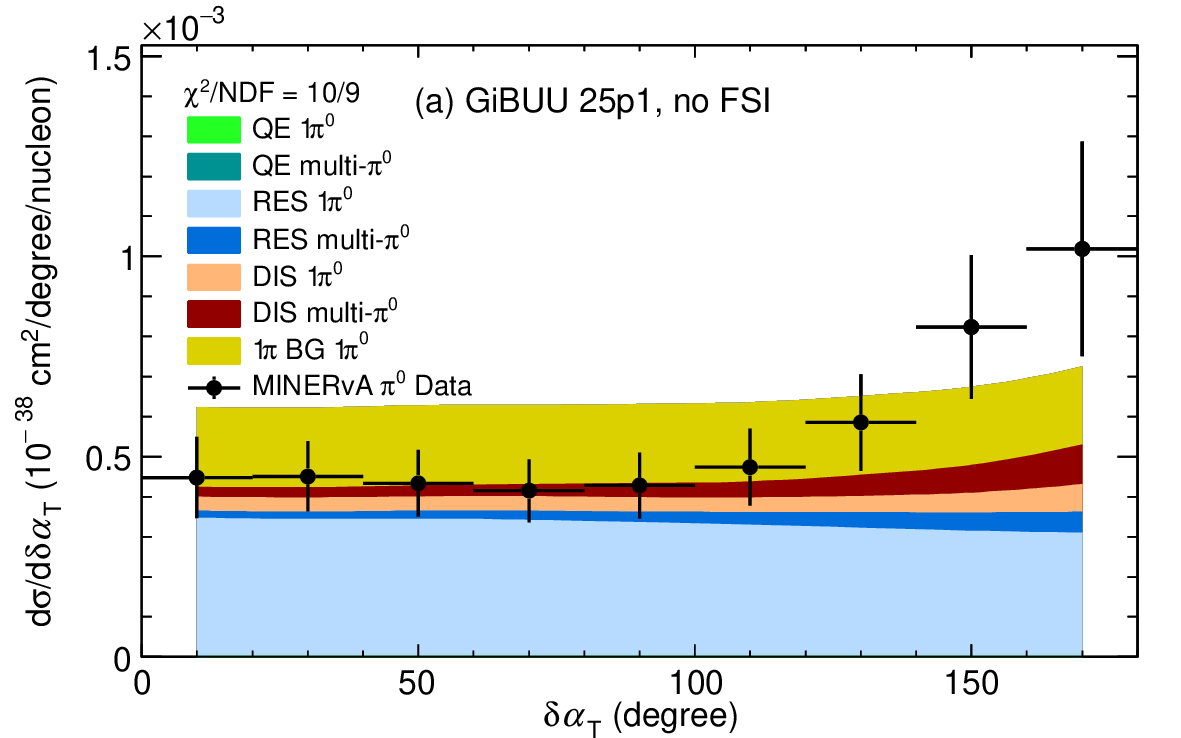}
    \includegraphics[width=0.4\textwidth]{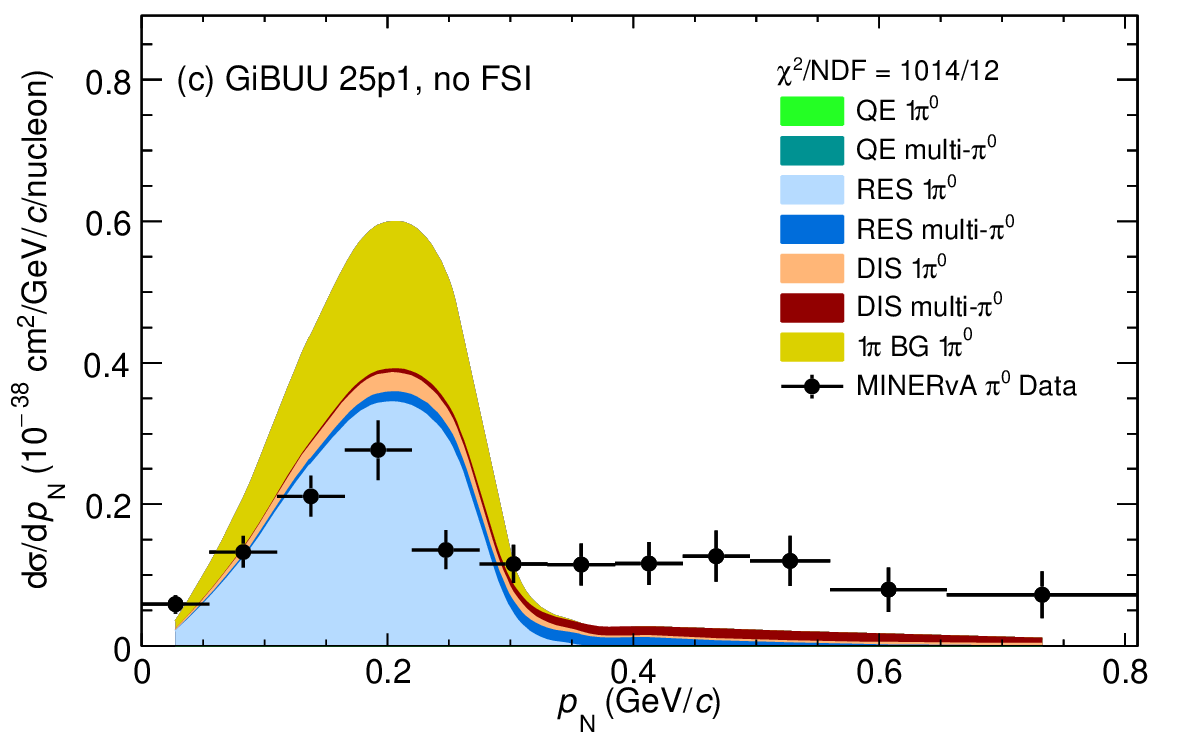}
    \caption{\gibuu CC$\pi^0$ TKI cross sections in (a) $\dat$ and (b) $\pn$~\cite{MINERvA:2020anu} calculated before FSI, the contribution from different interaction channels are decomposed.  } \label{fig:nofsi_pi0}
\end{figure}

This appendix section provides complementary information to the discussion in \cref{sec:tki}, showing results for events before final-state interactions (FSI) using similar variable definitions and signal selections. For the MINERvA CC0$\pi$ TKI case, as shown in \cref{fig:nofsi_0pi}, the final-state contribution is dominated by the QE and 2p2h channels, resulting in a uniform distribution in $\dat$ and a peak at the Fermi momentum in $\pn$. For the MINERvA CC$\pi^0$ TKI case, shown in \cref{fig:nofsi_pi0}, significant contributions from various channels are observed around the Fermi momentum in $\pn$, indicating that pion absorption during FSI is crucial for providing a good description of the data.

\section{$b$-dependence of $\pn$ prediction for MINERvA TKI}\label{app:bdep}
\begin{figure}[!htb]
    \centering
    \includegraphics[width=0.4\textwidth]{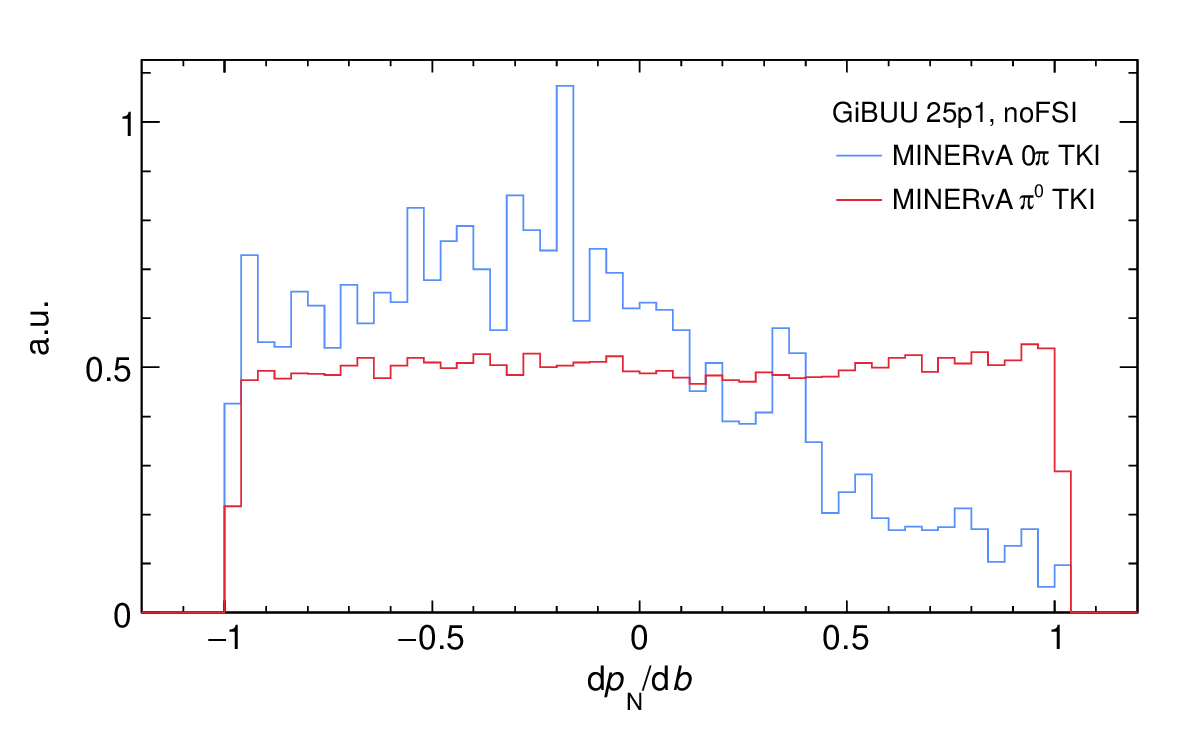}
    \caption{Distribution of $\frac{\mathrm{d} p_\text{N}}{\mathrm{d} b}$ computed from before-FSI samples for the MINERvA CC-$0\pi$ and CC-$\pi^0$ signal definitions. The distributions are normalized to unit area.}
    \label{fig:deriv_comparison}
\end{figure}

By applying the chain rule to Eqs.~\ref{eq:dpl} to \ref{eq:mstar}, the first-order derivative of $\pn$ with respect to the parameter $b$ is given by:
\begin{equation}
    \frac{\mathrm{d} p_\text{N}}{\mathrm{d} b} = \frac{\mathrm{d} p_\text{N}}{\mathrm{d} \delta p_L} \cdot \frac{\mathrm{d} \delta p_L}{\mathrm{d} M^*} \cdot \frac{\mathrm{d} M^*}{\mathrm{d} b} = -\frac{\delta p_\text{L} M^*}{p_\text{N} R}.
\end{equation}
Given that $|\delta p_\text{L}| < p_\text{N}$ and $M^* \sim R$, it is expected that the magnitude $\left| \frac{\mathrm{d} p_\text{N}}{\mathrm{d} b} \right|$ is bounded around 1. Substituting values from before-FSI event samples yields the distribution of $\frac{\mathrm{d} p_\text{N}}{\mathrm{d} b}$ shown in Fig.~\ref{fig:deriv_comparison}. The actual variation of $\pn$ with respect to variation in $\Delta b$ can be approximated as the distribution scaled by $\Delta b$.

\begin{figure}[!htb]
    \includegraphics[width=0.4\textwidth]{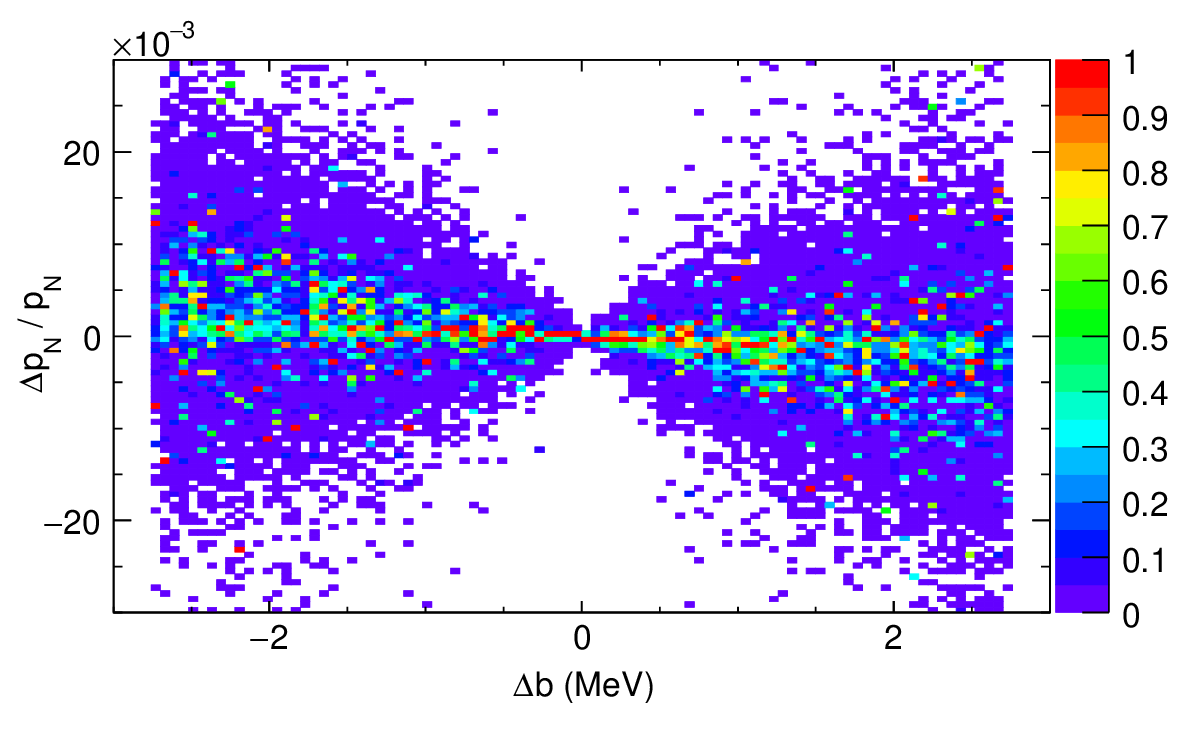}
    \includegraphics[width=0.4\textwidth]{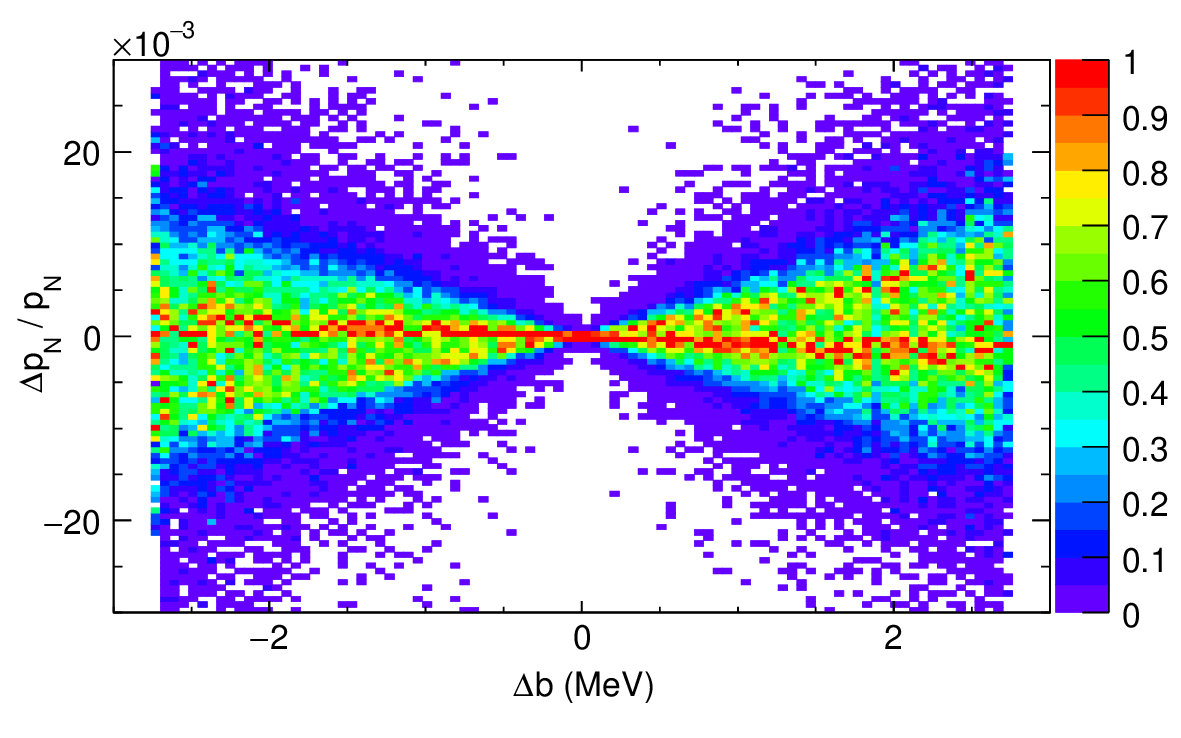}
    \caption{Relative change in $\pn$  as a function of the variation in $b$  for the \minerva 0$\pi$ (upper) and $\pi^0$ (lower)  TKI predictions. The maximum value for each $\Delta b$ slice is normalized to 1.}
    \label{fig:2dplot_bdep}
\end{figure}

\begin{figure}[!htb]
    \includegraphics[width=0.4\textwidth]{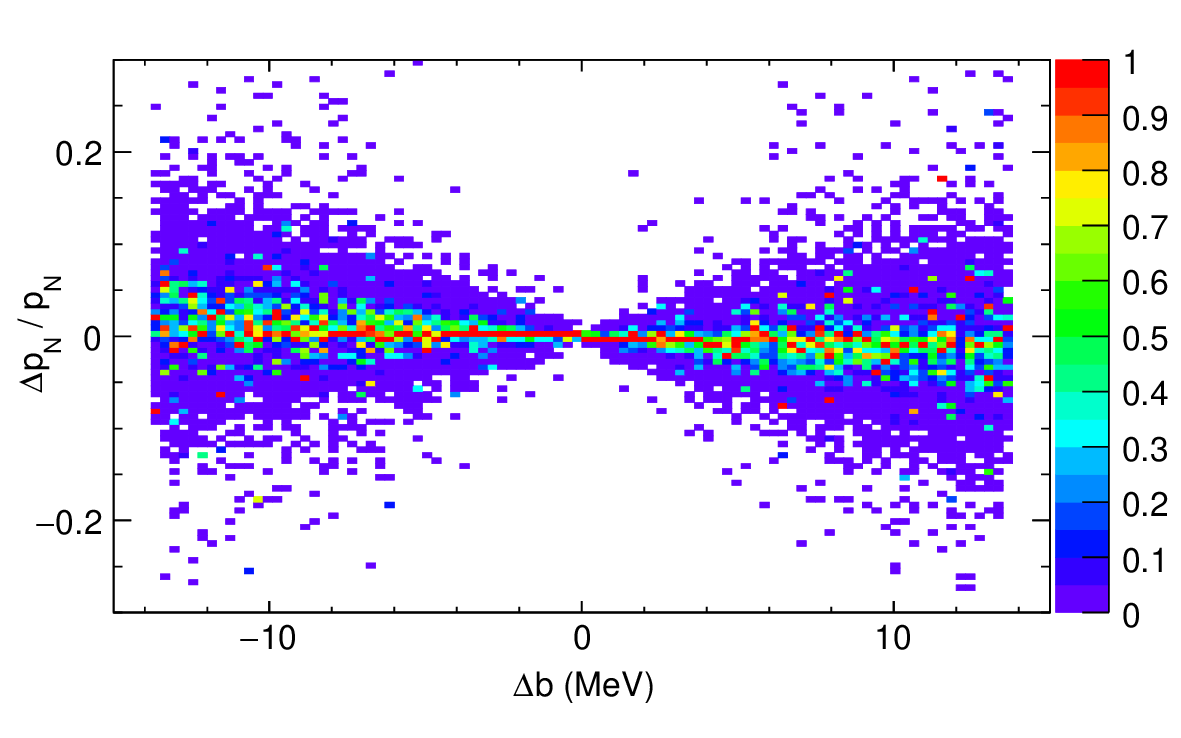}
    \includegraphics[width=0.4\textwidth]{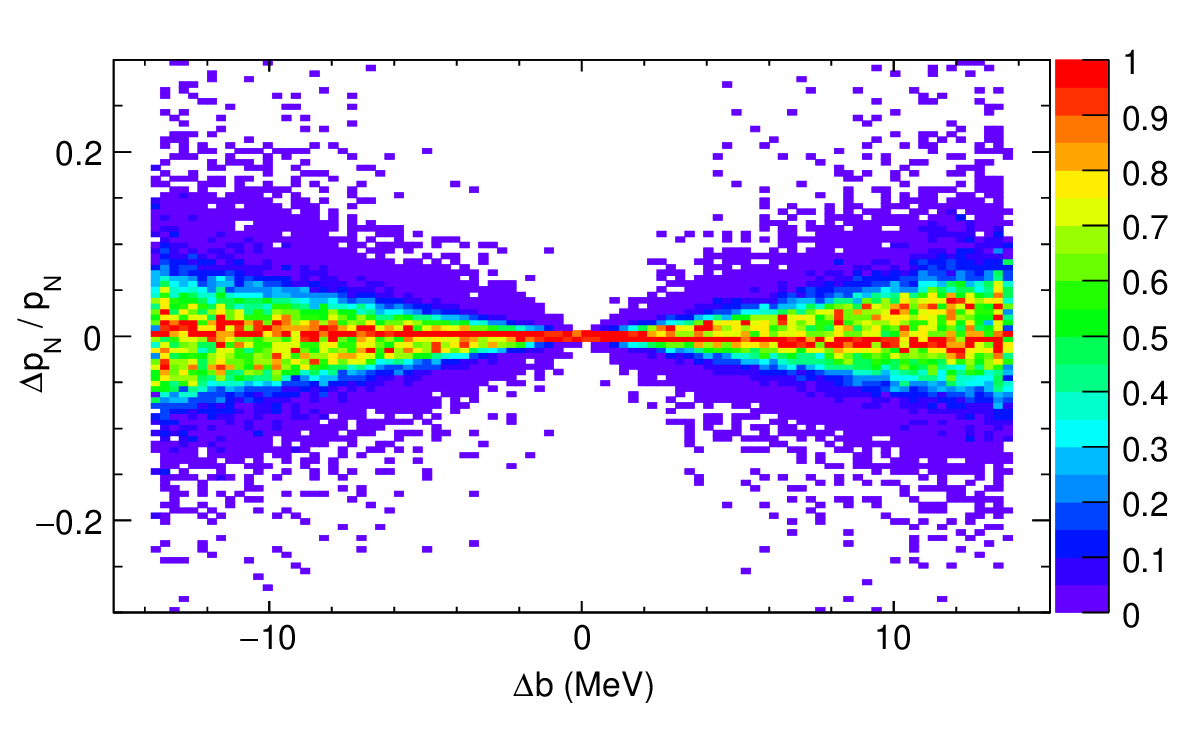}
    \caption{Similar to Fig.~\ref{fig:2dplot_bdep} but with a larger $\Delta b$ range.}
    \label{fig:2dplot_bdep_50}
\end{figure}

Figures~\ref{fig:2dplot_bdep} and ~\ref{fig:2dplot_bdep_50} show the event-by-event relative change of $\pn$ as a function of the variation of $b$  for the \minerva CC0$\pi$ and CC$\pi^0$ measurements~\cite{MINERvA:2018hba,MINERvA:2019ope,MINERvA:2020anu}. It can be seen that the variation of $b$ in a 10\% (50\%) range leads to a permil-(few-percent-)level change on  $\pn$.


\begin{figure}[!htb]
    \includegraphics[width=0.4\textwidth]{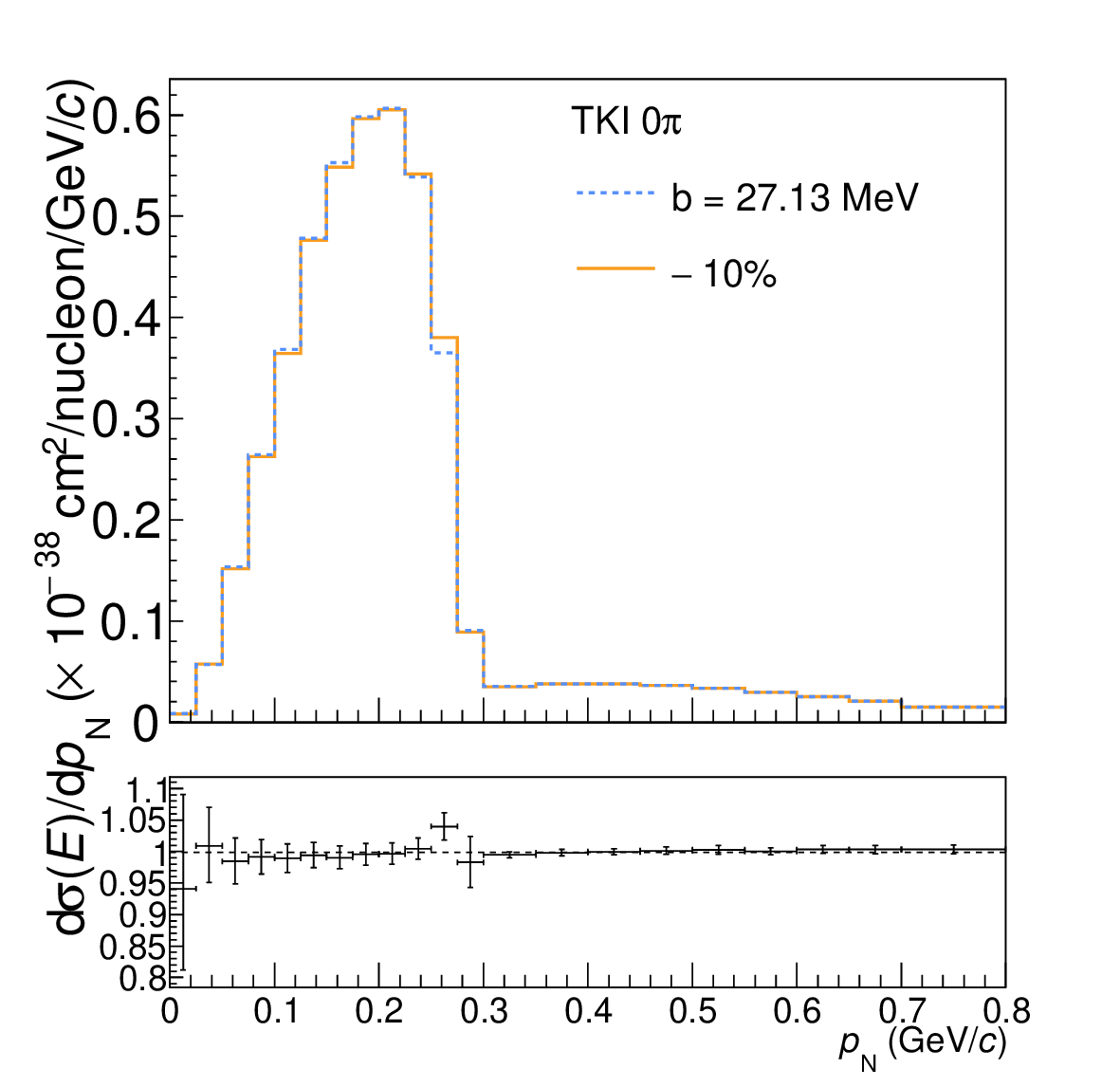}
    \includegraphics[width=0.4\textwidth]{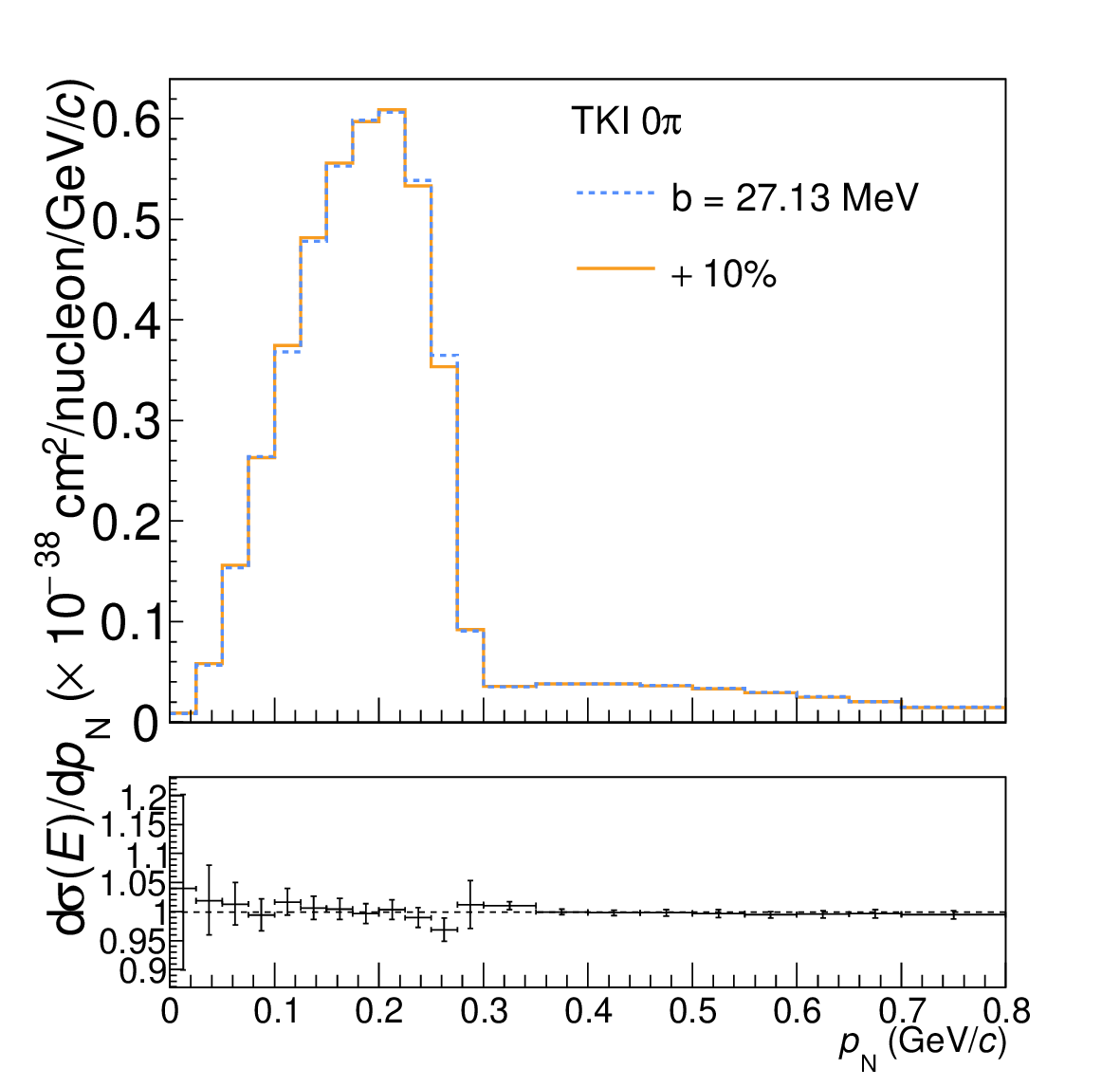}
    \caption{\gibuu CC0$\pi$ cross sections in $\pn$~\cite{MINERvA:2018hba,MINERvA:2019ope} calculated with different $b$ values. The bottom panel shows the ratio of the varied ($\pm10\%$) to the default case ($b=\SI{27.13}{MeV}$). All predictions are obtained before FSI.} \label{fig:appbdep}
\end{figure}

\begin{figure}[!htb]
    \includegraphics[width=0.4\textwidth]{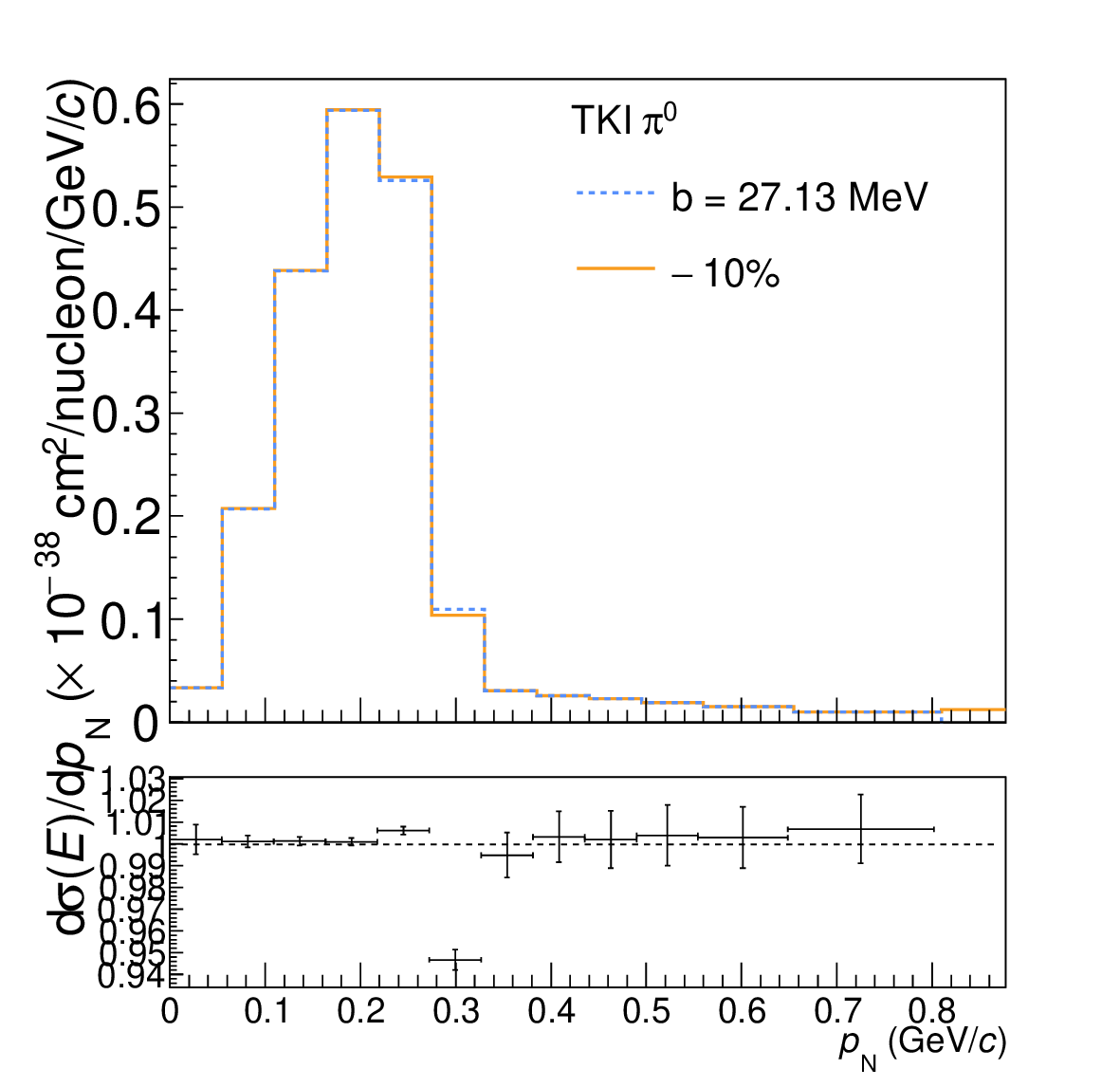}
    \includegraphics[width=0.4\textwidth]{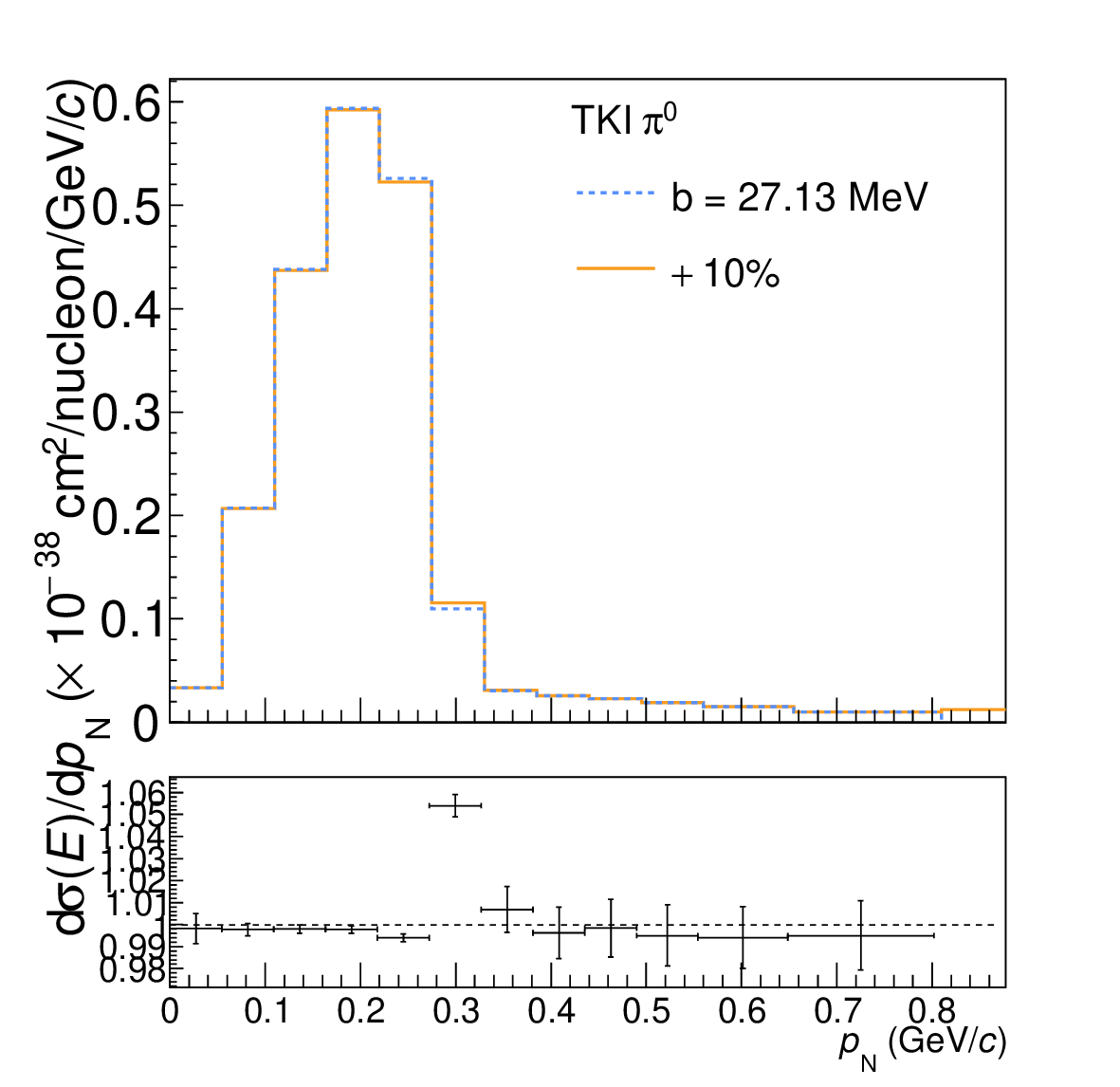}
    \caption{Similar to Fig.~\ref{fig:appbdep} but for \minerva CC$\pi^0$ TKI cross sections in $\pn$~\cite{MINERvA:2020anu}.} \label{fig:appbdep_pi0}
\end{figure}

Figures~\ref{fig:appbdep} and~\ref{fig:appbdep_pi0} show the $b$-dependence of $\pn$ cross sections. The $b$ parameter in Eq.~\ref{eq:mstar} is varied by 10\% from the default value of $\SI{27.13}{MeV}$. The ratios of the distributions are displayed in the lower panels. The impact of $b$ on the $\pn$ distribution is negligible.


\section{\minerva inclusive cross section decomposed by interaction channels}\label{app:xsec_decompose}
\begin{figure}
    \includegraphics[width=0.4\textwidth]{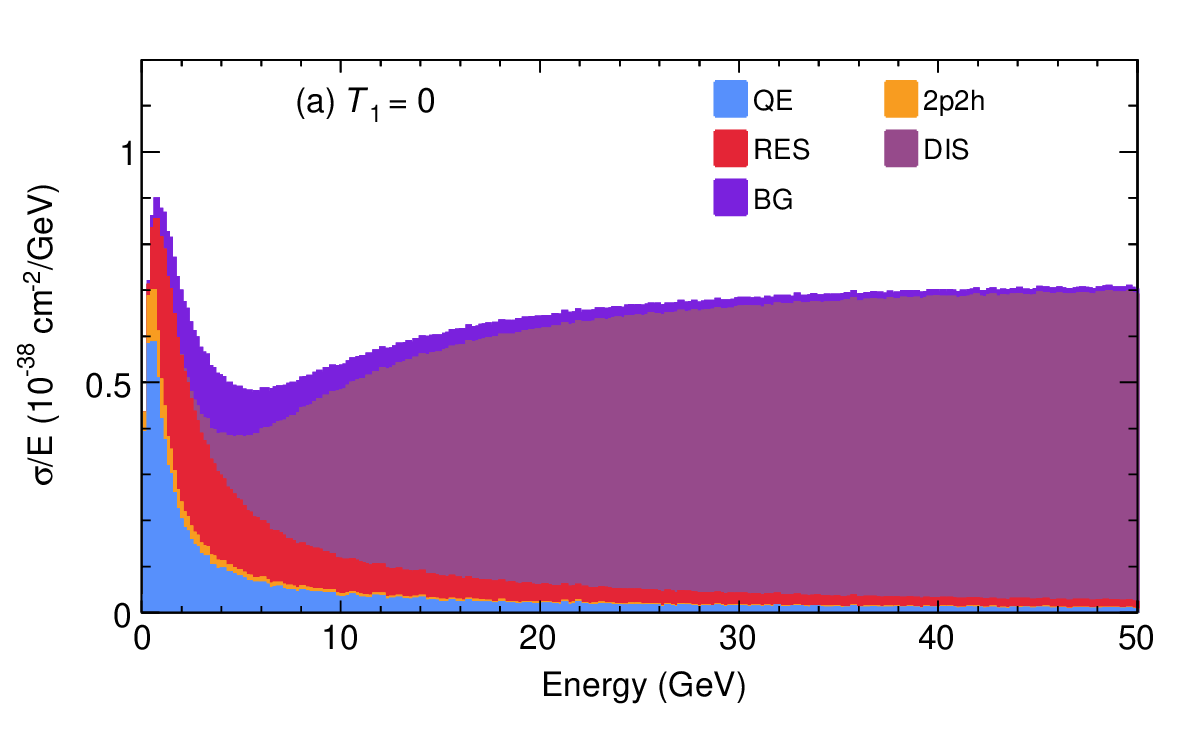}
    \includegraphics[width=0.4\textwidth]{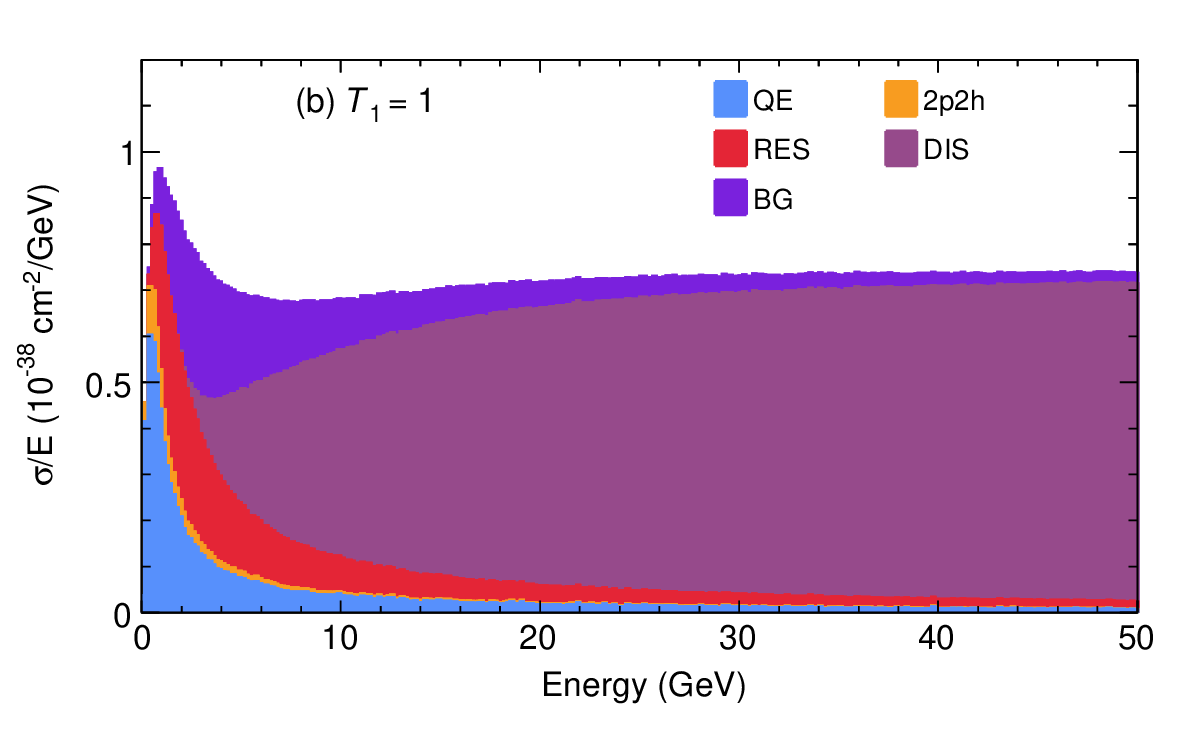}
    \caption{\gibuu inclusive cross sections decomposed by interaction channel with (a) $\Toneb=0$ and (b) $\Toneb=1$, see also Fig.~\ref{fig:different_t2p2h}.} \label{fig:inclusive_decom}
\end{figure}

For completeness, we provide in this appendix the decomposition of the \minerva inclusive cross section shown in Fig.~\ref{fig:different_t2p2h} in \cref{fig:inclusive_decom}, to demonstrate the contributions from different interaction channels and how they vary with $\Toneb$.

\section{\minerva, T2K, and \microboone CC0$\pi$ TKI measurements}\label{app:minervacc0pi}

Results in this appendix give for completeness the decomposition into the various production mechanisms for the observables already discussed in the main text.

\begin{figure}[!htb]
    \centering
    \includegraphics[width=\figwid]{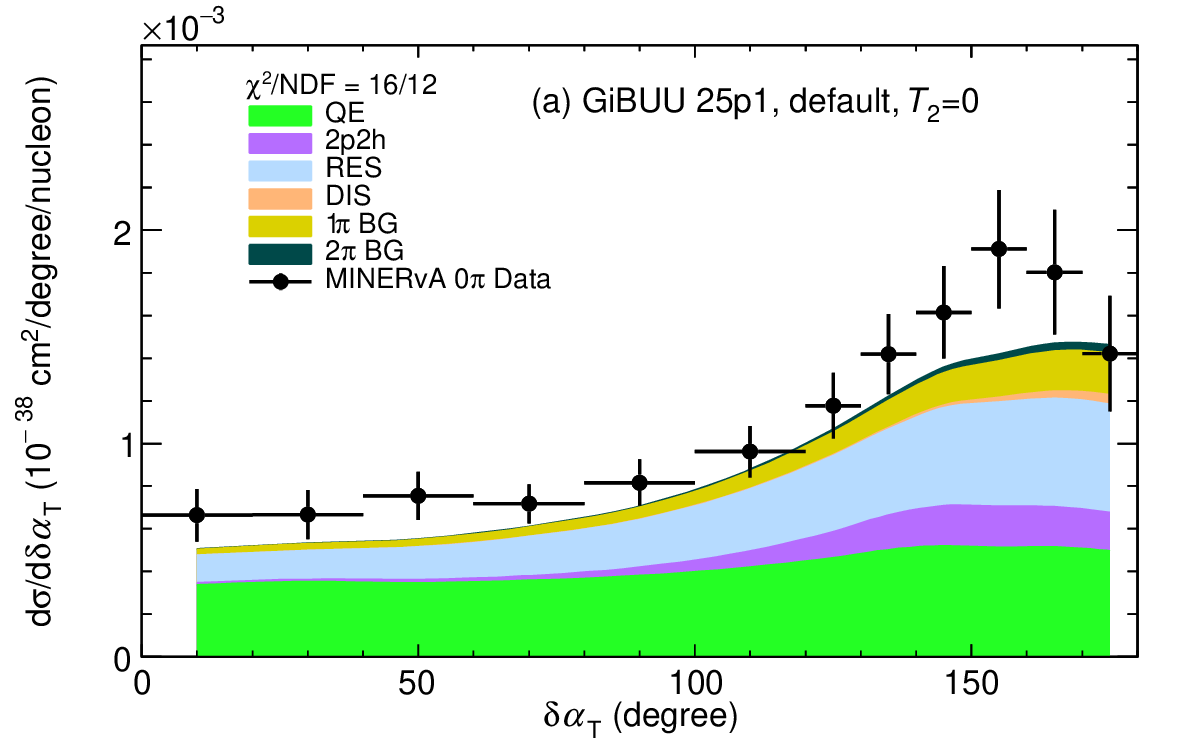}
    \includegraphics[width=\figwid]{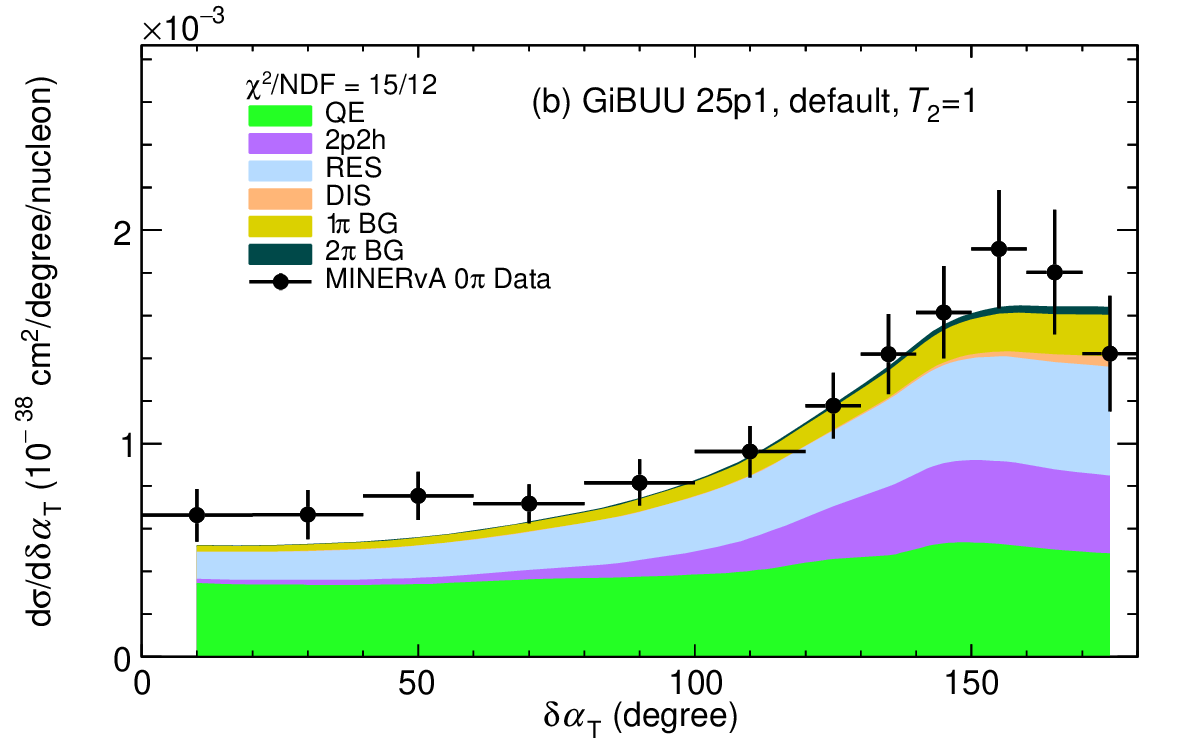}
    \caption{\minerva CC0$\pi$ TKI cross sections in $\dat$~\cite{MINERvA:2018hba,MINERvA:2019ope} compared to  \gibuu  predictions with the default configuration and (a) $\Ttpth=0$  and (b) $\Ttpth=1$. }
    \label{fig:m0pit0}
\end{figure}

\begin{figure}[!htb]
    \centering
    \includegraphics[width=\figwid]{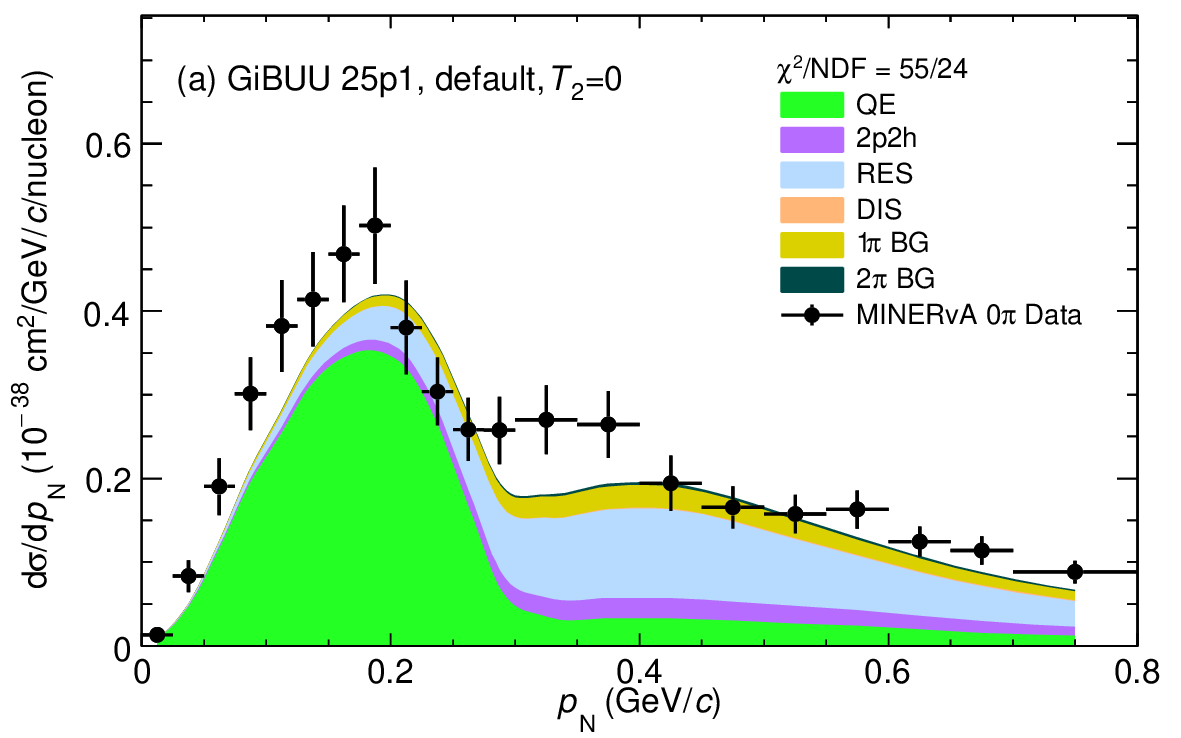}
    \includegraphics[width=\figwid]{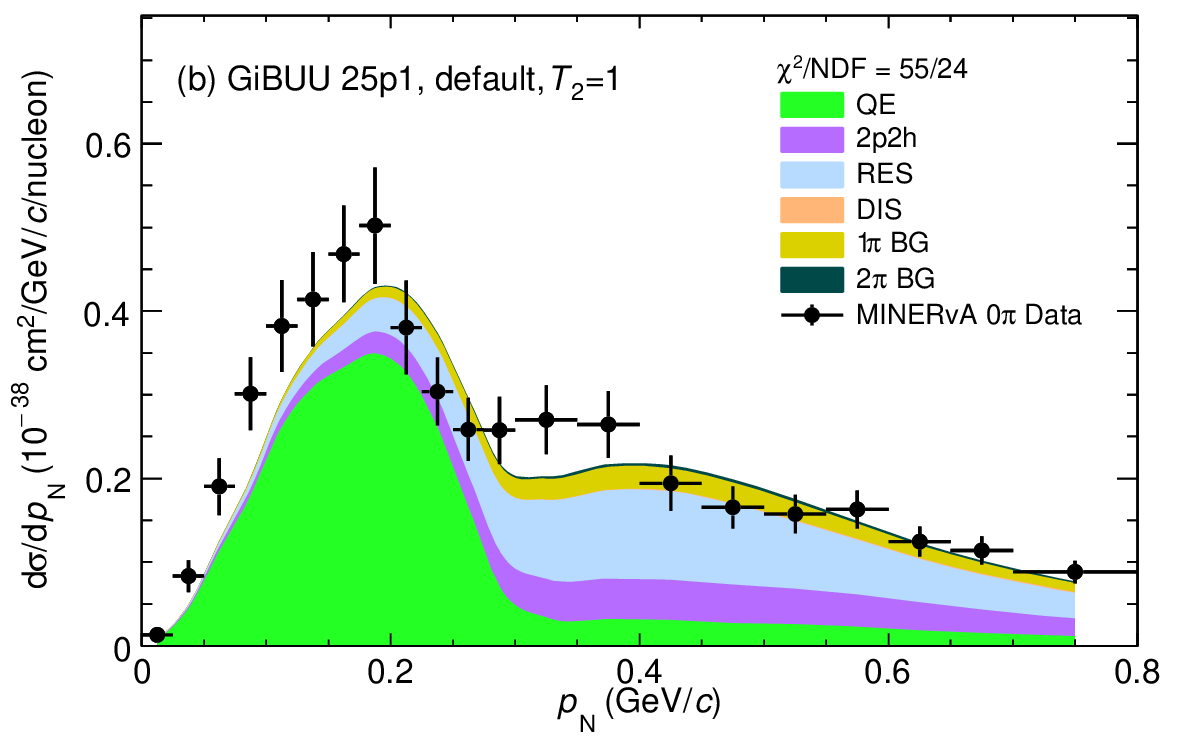}
    \caption{Similar to Fig.~\ref{fig:m0pit0} but for $\pn$.}
    \label{fig:m0pit1}
\end{figure}

\begin{figure}[!htb]
    \centering
    \includegraphics[width=\figwid]{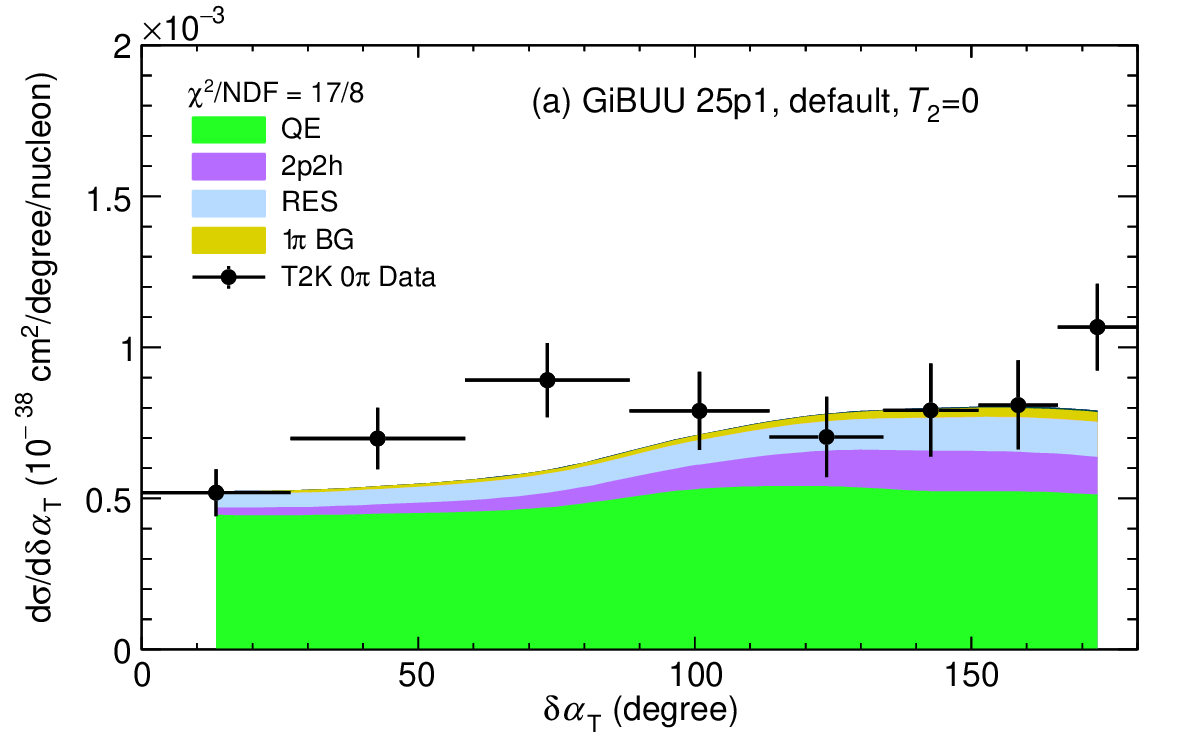}
    \includegraphics[width=\figwid]{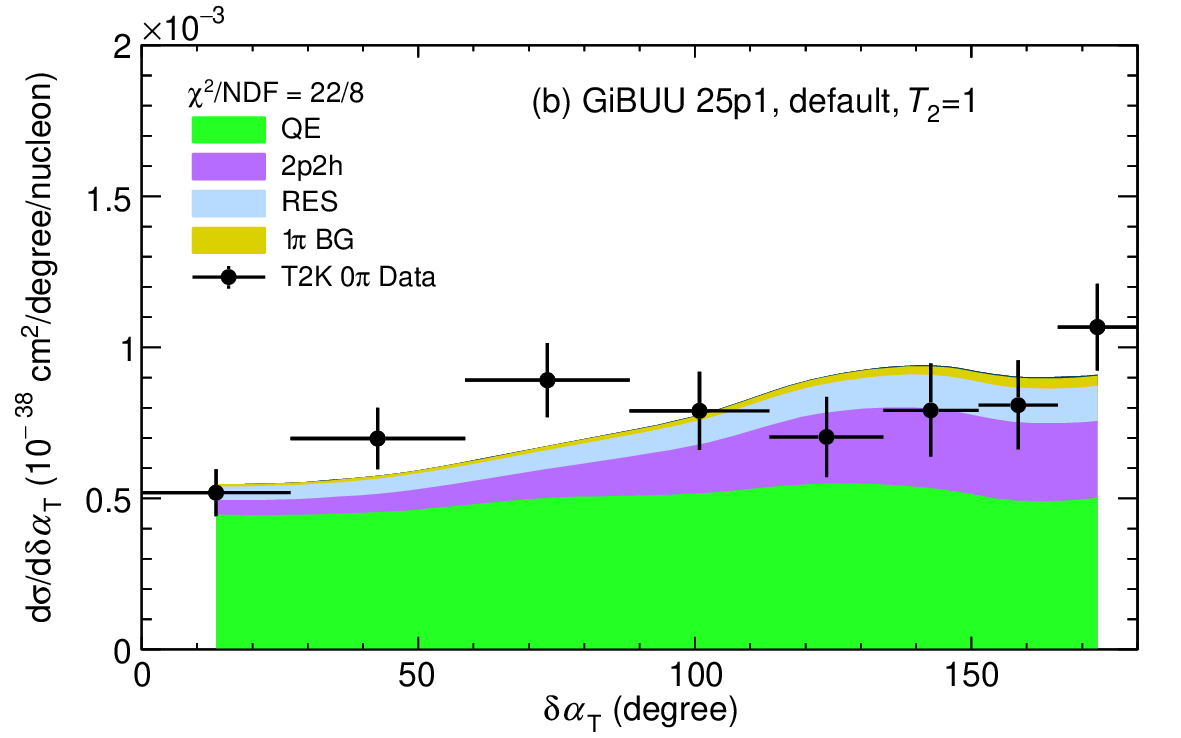}
    \caption{T2K CC0$\pi$ TKI cross sections in $\dat$~\cite{T2K:2018rnz} compared to \gibuu  predictions with the default configuration and  (a) $\Ttpth=0$ and (b) $\Ttpth=1$.  }
    \label{fig:T2K_dalphat_0pi}
\end{figure}

\begin{figure}[!htb]
    \centering
    \includegraphics[width=\figwid]{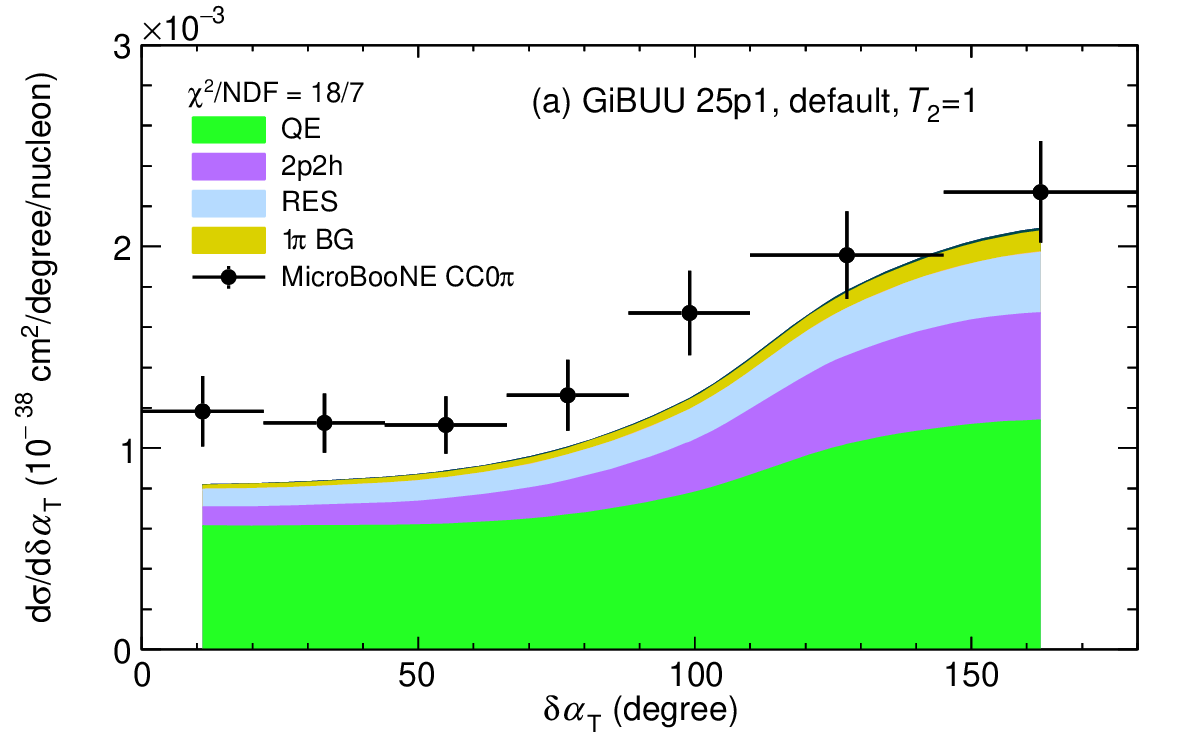}
    \includegraphics[width=\figwid]{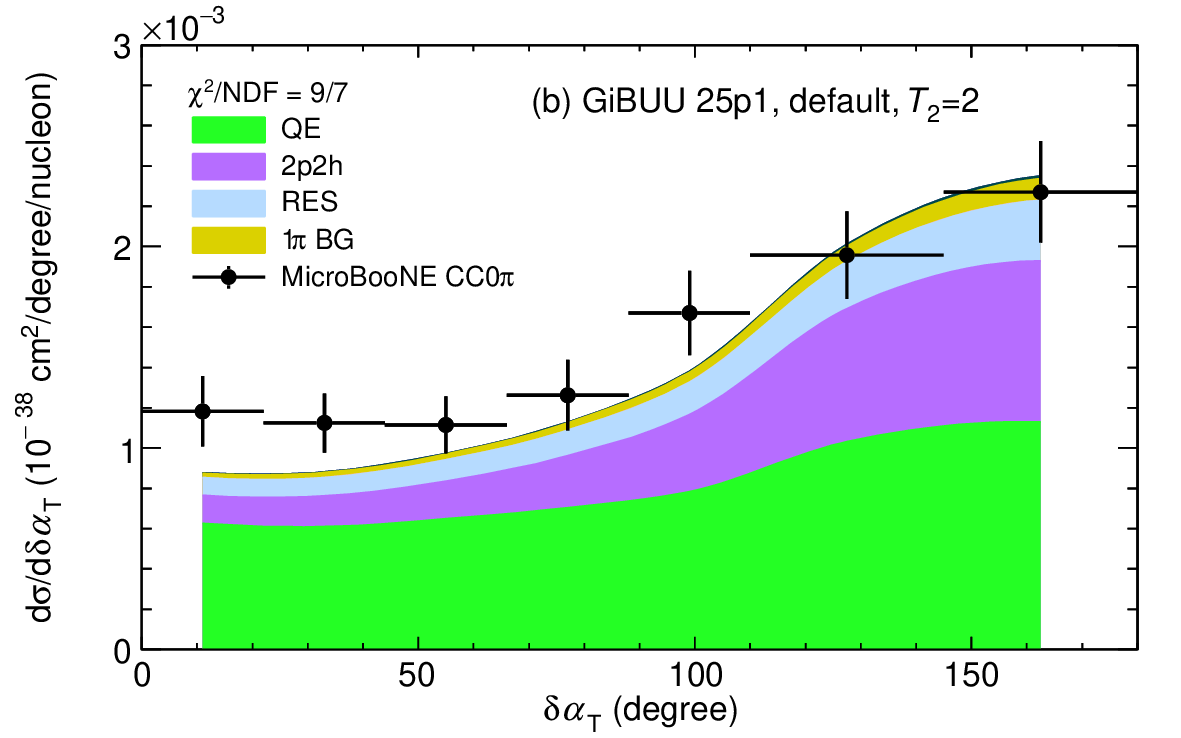}
    \caption{MicroBooNE CC0$\pi$ TKI cross sections in $\dat$~\cite{MicroBooNE:2023cmw} compared to \gibuu  predictions with the default configuration and  (a) $\Ttpth=1$ and (b) $\Ttpth=2$.  }
    \label{fig:ub_dalphat_0pi}
\end{figure}

Figures~\ref{fig:m0pit0}--\ref{fig:ub_dalphat_0pi} compare the \minerva~\cite{MINERvA:2018hba,MINERvA:2019ope},  T2K~\cite{T2K:2018rnz}, and \microboone~\cite{MicroBooNE:2023cmw} TKI cross sections with the \gibuu predictions.

\clearpage %

\end{document}